\documentclass[onecolumn]{emulateapj}
\lefthead{PEPPER and BURKE} \righthead{VARIABLES IN NGC 1245}%\documentclass[preprint2]{aastex}
\usepackage{graphicx}
\begin{document}
\def\ch{{\bf changed}\ }

\title{Survey for Transiting Extrasolar Planets in Stellar Systems IV: Variables in the Field of NGC 1245}

\author{Joshua Pepper and Chris Burke}
\affil{Department of Astronomy, Ohio State University, Columbus, Ohio 43210}
\email{pepper@, cjburke@astronomy.ohio-state.edu}

\begin{abstract}

The Survey for Transiting Extrasolar Planets in Stellar Systems
(STEPSS) project is a search for planetary transits in open clusters.
In this paper, we analyze the STEPSS observations of the open cluster
NGC 1245 to determine the variable star content of the cluster.  Out
of 6787 stars observed with $V < 22$, of which $\sim 870$ are cluster
members, we find 14 stars with clear intrinsic variability that are
potential cluster members, and 29 clear variables that are not cluster
members.  None of these variables have been previously identified.  
We present light curves, finding charts, and
stellar/photometric data on these variable objects.  Several of the
interacting binaries have estimated distances consistent with the
cluster distance determined from isochrone fits to the color magnitude
diagram.  Four stars at the main sequence turnoff of the cluster
have light curves consistent with $\gamma$ Doradus variability.  If these
$\gamma$ Doradus candidates are confirmed, they represent the oldest and
coolest members of this class of variable discovered to date.

\end{abstract}

\section{INTRODUCTION}

The dramatic increase in the number of surveys for planetary transits
over the last five years has resulted in a wealth of high-precision
stellar photometry \citep{horne03}.  Although primarily conducted to
detect planetary transits, such surveys also produce data sets that
allow for intensive investigation of stellar variability.  Especially
useful are surveys of clusters, which can be used to characterize the
variability content of the cluster.  Detections of eclipsing binaries
can provide a check on the cluster's distance, as well as helping to
characterize the mass-radius and mass-luminosity relationships.

Many surveys of clusters have already published data sets of
high-precision photometry, finding low-amplitude variables, eclipsing
binaries, and other pulsating stars.  The PISCES (Planets in Stellar
Clusters Extensive Search) project has observed clusters NGC 2158 and
NGC 6791 \citep{mo02,mo04,mo05}.  The EXPLORE/OC (Extrasolar Planet
Occultation Research/Open Clusters) project has observed clusters NGC
2660 and NGC 6208 \citep{vonb05}.  Other projects have observed or are
observing NGC 7789, 6819, and 6940 \citep{street03,bram05}, NGC 6633
\citep{hidas05}, and NGC 6705 \citep{harg04}.

The Survey for Transiting Extrasolar Planets in Stellar Systems
(STEPSS) \citep{burke02} concentrates on searching open clusters for
planetary transits.  Searching for transits in clusters is
advantageous as clusters have known metallicities, ages, and stellar
densities, and are therefore good laboratories for determining the
distribution of planets without varying those stellar and
environmental properties.  Here we present an analysis of the
observations of NGC 1245 to catalog the stellar variables in the
cluster.

\section{CLUSTER PARAMETERS AND OBSERVATIONS} \label{sec:observations}

See \citet{burke04} for a full discussion of the parameters of the
cluster.  NGC 1245 is a rich open cluster with an age of $1.04 \pm 0.09$ Gyr.
It has a metallicity slightly less than solar, [Fe/H] $= -0.05 \pm
0.08$.  The cluster is at a distance of $2.8 \pm 0.2$ kpc, with
distance modulus of $(m - M)_0 = 12.27 \pm 0.12$.  The core radius is
$r_c = 3.10 \pm 0.52$ arcmin ($2.57 \pm 0.43$ pc).  The total cluster
mass is $M = 2700 \pm 600 M_{\odot}$.  All errors given above are
systematic, and are larger than the statistical errors.  

The STEPSS project observed NGC 1245 over the course of 19 nights in
October and November 2001.  The observations were obtained with the
MDM 8K mosaic imager on the MDM 2.4m Hiltner telescope yielding a
$26\arcmin \times 26\arcmin$ field of view with $0.36\arcsec$
resolution per pixel.  In total, the dataset consists of 936 $I$-band
images with typical exposure times of 300 s.  For the color magnitude diagram (CMD), several
supplementary $B$ and $V$ band images were also obtained.  None of the
nights were photometric, and so a second series of observations was
taken in February 2002 to calibrate the cluster photometry.

\section{LIGHT CURVES AND STELLAR PROPERTIES} \label{sec:membership}

We follow the procedure outlined in \citet{burke05} to generate light
curves, which develops a new method for
differential photometry given the demands of detecting transiting extrasolar planets
($<0.01$ mag precision).  The main features of the method include
allowing each star to have a unique comparison ensemble, minimization
of the light curve standard deviation as a figure of merit, and full
automation.  Applying this method yields light curves that have sub-1\% precision for $V<18$.

\begin{figure}
\epsscale{1.0}
\plotone{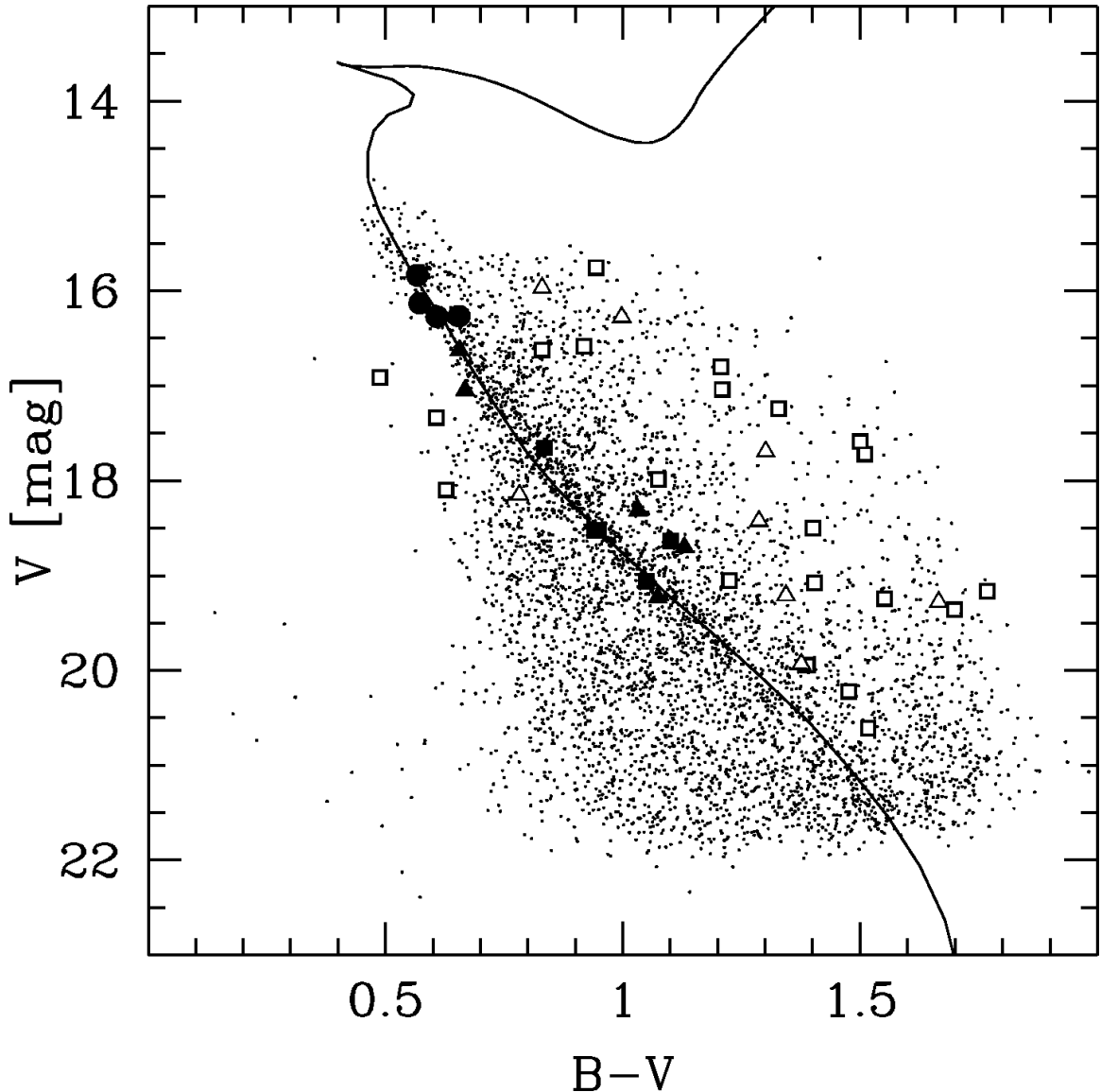}
\caption{$V$, $\bv$ CMD for NGC 1245.  Best-fit isochrone to the cluster main 
sequence ({\it solid line}).  Filled shapes represent cluster variables, while open shapes represent field variables.  The triangles show binary variables (see \S\ref{sec:binaries}), the circles show $\gamma$ Doradus 
candidates (see \S\ref{sec:gamdor}), and the squares represent unclassified variables. }
\label{fig:cmd}
\end{figure}

Of the 6787 stars identified in the survey's field of view, we
estimate $\sim 870$ belong to the cluster.  This estimate comes from
scaling and subtracting star counts from a control field on the
outskirts of our field of view.  Figure \ref{fig:cmd} shows the $V$,
$\bv$ CMD of the cluster field along with the best-fit isochrone from
\citet{burke04}, derived from the theoretical isochrones of
\citet{yi01}.  Given the cluster age, metallicity, distance, and
reddening, the best-fit isochrone transforms an observed apparent
magnitude into the stellar properties assuming the star is a cluster
member.  Using the stellar mass as the independent variable, we
determine the stellar mass which minimizes the $\chi^{2}$ distance
between the apparent magnitude and the best-fit isochrone magnitudes
in the $BVI$ passbands.

A caveat for assigning stellar properties based on the CMD is the
assumption that the detected stellar light arises from a single
main-sequence cluster member.  The stellar properties contain
significant systematic errors if the object does not belong to the
cluster or if a portion of the stellar light arises from an unresolved
stellar companion.  The physical properties for the variable sources
detected in this study should be regarded with extra caution
considering that most sources of variability result from the binary
nature of an object \citep{sterken96} and thus have a large
probability for significant stellar companion contamination.

\section{VARIABLE STAR SELECTION} \label{sec:varselect}

One of the great challenges in determining variability with high
precision photometry is the difficulty of distinguishing intrinsic
variability from noise and systematic errors.  Common classes of
variable stars demonstrate both large amplitude variations and
periodic behavior, making them both easily detected and easily 
characterized.  However, with the advent of surveys which observe thousands
of stars with millimagnitude photometry, robustly detecting variations
at low noise levels requires a variety of statistical selection
criteria.  Some surveys, such as ASAS, the All Sky Automated Survey
\citep{eyer05}, and OGLE, the Optical Gravitational Lensing Experiment
\citep{miz02}, have developed automated classification pipelines.
Many surveys, though, use a few simple statistical cuts to identify
variable stars independent on the source or type of variability.

Our variability selection routine has three stages.  First, we
eliminate outlying measurements, and require a minimum number of
remaining measurements.  Second, we compute three different statistics
that describe the behavior of the lightcurve: the root-mean-square
(RMS), the Stetson $J$ statistic \citep{stet96}, and the AoV
periodicity statistic \citet{sc96}.  We apply cuts on each of those
values to select variable candidates.  Third, we visually inspect the
lightcurves and images of the variable candidates to eliminate false
positive candidates with variability due to blending effects from
nearby objects or detector defects.  In \S\ref{sec:finalcuts}, we
describe two tiers of statistical cuts.  The first, more stringent cut does
not contain any false positives.  A second, less stringent, cut with higher 
sensitivity contains lower amplitude variables, but also contains $<$ 7\% false
positives, which are later eliminated through visual inspection.

\subsection{Eliminating Outlying Points\label{sec:outlying}}

The first stage of the analysis removes statistically outlying photometric 
measurements.  Outliers are identified in the following manner:  For each 
star we calculate the reduced chi-square light curve variability,
\begin{equation} \label{equ:chi}
\chi^2 = \frac{1}{n-1} \sum_{k=1}^{n} \frac{(m_k - \mu)^2}{\sigma_k^2},
\end{equation}
where the sum is over $n$ observations, $m_k$, with error $\sigma_k$,
and $\mu$ is the weighted average magnitude of the light curve,
\begin{equation} \label{equ:wmu}
\mu = \frac{\sum_{k=1}^{n} m_k/\sigma_k^2}{\sum_{k=1}^{n} 1/\sigma_k^2} \, .
\end{equation}
The $\chi^2$ statistic measures the degree to which a constant light
curve model approximates the light curve data within the context of
the measurement errors.  Light curves with intrinsic variability
result in high $\chi^{2}$ values, denoting that a constant light curve
model poorly represents the data.  Before eliminating outlying
measurements, we scale the errors by setting $\sigma_k^{'2} =
\sigma_k^2 \chi^2$, thus enforcing $\chi^{2}=1$.  The error scaling
allows the Gaussian noise, systematic error, or intrinsic variability
(whichever noise source dominates the scatter in the light curve) to
set the scale for judging whether a measurement is a statistical
outlier.  This procedure can eliminate intrinsic variability
that affects a small number of the measurements.  This
hampers our ability to detect large amplitude
eclipsing binaries when only a single eclipse occurred
during the observations.  However, given our quantitative
selection criteria and documented analysis techniques,
these biases can easily be accounted for with a more
detailed inspection of the data.

For each observation, we compute the deviation, $\delta_k = |m_k -
\mu|/\sigma_k^{'}$.  All photometric measurements with $\delta_k > 3$
are eliminated.  We perform three iterations of the above procedure
for each light curve.  The light curves with $<$750 measurements
remaining typically result from false stellar detections near the
cores of bright isolated stars and are discarded.  We also eliminate
light curves with RMS $< 0.3$ mag, since no light curves with higher
scatter show any kind of coherent intrinsic variability.

\subsection{The Stetson $J$ Statistic}

The Stetson $J$ statistic \citep{stet96} provides our primary measure of
the coherent intrinsic light curve variability.  The 
statistic measures variability by weighting photometric variations
that are correlated in time.  A light curve with a steady change in
brightness over a period of time results in a higher $J$ value than a
curve with Gaussian noise, even though both curves may have equal
values for $\chi^2$.  The statistic is defined by
\begin{equation} \label{equ:J}
J = \frac{\sum_{k=1}^{n} w_k \,\, {\rm sgn}(P_k) \sqrt{|P_k|}}{\sum_{k=1}^{n} w_k},
\end{equation}
where the observer defines $n$ pairs of observations with weights
$w_k$.  We define observations as a pair when they are separated by
less than 0.02 days.  $P_k$ is the product of the normalized residuals
of two observations $i$ and $j$, constituting the $k$th pair, such
that
\begin{equation} \label{equ:Pk}
P_k = \left\{ \begin{array}{ll} 
\delta_{i(k)}\delta_{j(k)}, & \mbox{if $i(k) \ne j(k)$} \\
\delta_{i(k)}^2 - 1, & \mbox{if $i(k) = j(k)$} 
\end{array} \right .
\end{equation}
and $\delta$ is the ``relative error'', defined as
\begin{equation} \label{equ:delta}
\delta = \sqrt{\frac{n}{n-1}} \frac{m - \mu}{\sigma}, 
\end{equation}
where $m$ is the apparent magnitude of the star, $\sigma$ is the error
in the magnitude, and $\mu$ is the weighted mean magnitude of the
star.  Following the choices of \citet{kal98}, we set $w_k = 1$ for
pairs of observations ( $i(k) \ne j(k)$ ) and $w_k = 0.25$ for single
observations ( $i(k) = j(k)$ ), and we also multiply the final
quality $J$ by $\sum_{k=1}^{n} w_k / w_{max}$, where $w_{max}$ is the
total weight the star would have if it were measured on all images.

Of the stars we observe, the Stetson statistic varies from $J \ll 1$
for nonvariable sources, to $1 < J < 10$ for obvious variables.  For
objects with $J \approx 1$, it can be difficult to demarcate a
specific cut in $J$ to specify variability.  We describe our choices
for cuts on $J$ in \S \ref{sec:finalcuts}.

\subsection{Periodicity Analysis}

To select periodic variations, we apply the Multiharmonic Analysis of
Variances (ANOVA) period-search algorithm described by \citet{sc96}.
Given a range of periods, the ANOVA algorithm produces a likelihood
statistic, $0 \leq AoV \leq 1$, that returns the quality of fit for
orthonormal polynomials to the light curve variability.  Since the
observations occurred over a duration of 19 days, we run the ANOVA
algorithm over a period range of 0.1 days to 14 days, which we believe
to be the variability periods that we could reasonably detect.

One problem with applying the ANOVA algorithm to ground-based
observations is the appearance of significant light curve variability
on the diurnal period and its aliases.  The rise and set of the field
of observation and the corresponding secular changing atmospheric path
length and mechanical stresses on the telescope impart systematic
errors in the light curve.  Our light curves display trends in the
light curve variability as a function of seeing.  In
\S\ref{sec:membership} we correct for these trends, but trends above
the expected Gaussian noise remain.  We find signs of aliasing
primarily at periods of 1/3, 1/2, 2/3, 1, 2, 3, and 4 days.  Relying
on the ANOVA statistic alone results in many detections of low
amplitude ($<$ 0.01 mag) variables with periods at the diurnal period
and its aliases.  However, in combination with the Stetson $J$
statistic (see \S\ref{sec:finalcuts}) these alias variables do not
make it into the final variable object samples.

\subsection{Final Cuts and Visual Inspection} \label{sec:finalcuts}

There are several systematic effects that complicate the attempt to
impose categorical cuts in $J$ and $AoV$ that result in a sample 
that includes all probable intrinsic variables and
zero false detections.  Systematic errors in the light curve, such as
blended objects, edge effects, detector defects, and detector saturation
can artificially inflate $J$ and $AoV$ for non-variable sources.
Furthermore, such effects, along with poor photometry, can in some
cases lead to lower $J$ and $AoV$ values for true variables, compared
to perfect photometry.  We conclude that the most useful evaluation of
the variability in our sample is a two-tier selection, in which we
identify stars with robust variability and zero false detections, and
then a second tier of candidate variables with a less stringent
selection and resulting in sample with some low contamination due to false
detections.

Our first tier consists of stringent cuts on the $J$ and $AoV$
statistical values.  We find that cuts of $J>1.2$ and $AoV>0.65$ yield
an ensemble of 23 stars which are all clearly intrinsic variables.
The first tier has zero contamination due to false positives.
However, it misses a number of other clear variables that vary with
lower amplitudes.  We then conduct a visual inspection of the images
of the remaining sources, and flag those that are near detector
defects such as bad columns, diffraction spikes, bleed trails, etc.  A
second tier cut of $J > 0.75$ and $AoV > 0.55$ includes most stars in
our sample that have coherently varying light curves and includes few
false positives.  This second cut contains 44 variable candidates,
including all the first tier objects.  Visual inspection of the
lightcurves and images of these stars enables us to eliminate three of
the second tier stars due to detection effects, yielding a
contamination rate of $\sim 7\%$.  Based on a visual inspection of
light curves, the contamination rate climbs rapidly for lower
statistical cuts.  Upon visual inspection of stars that do not make
the cuts, we do find two more field stars that show coherent, periodic
variability at low amplitudes.  We include these two stars in our
field star sample.

\begin{figure}
\epsscale{1.0}
\plotone{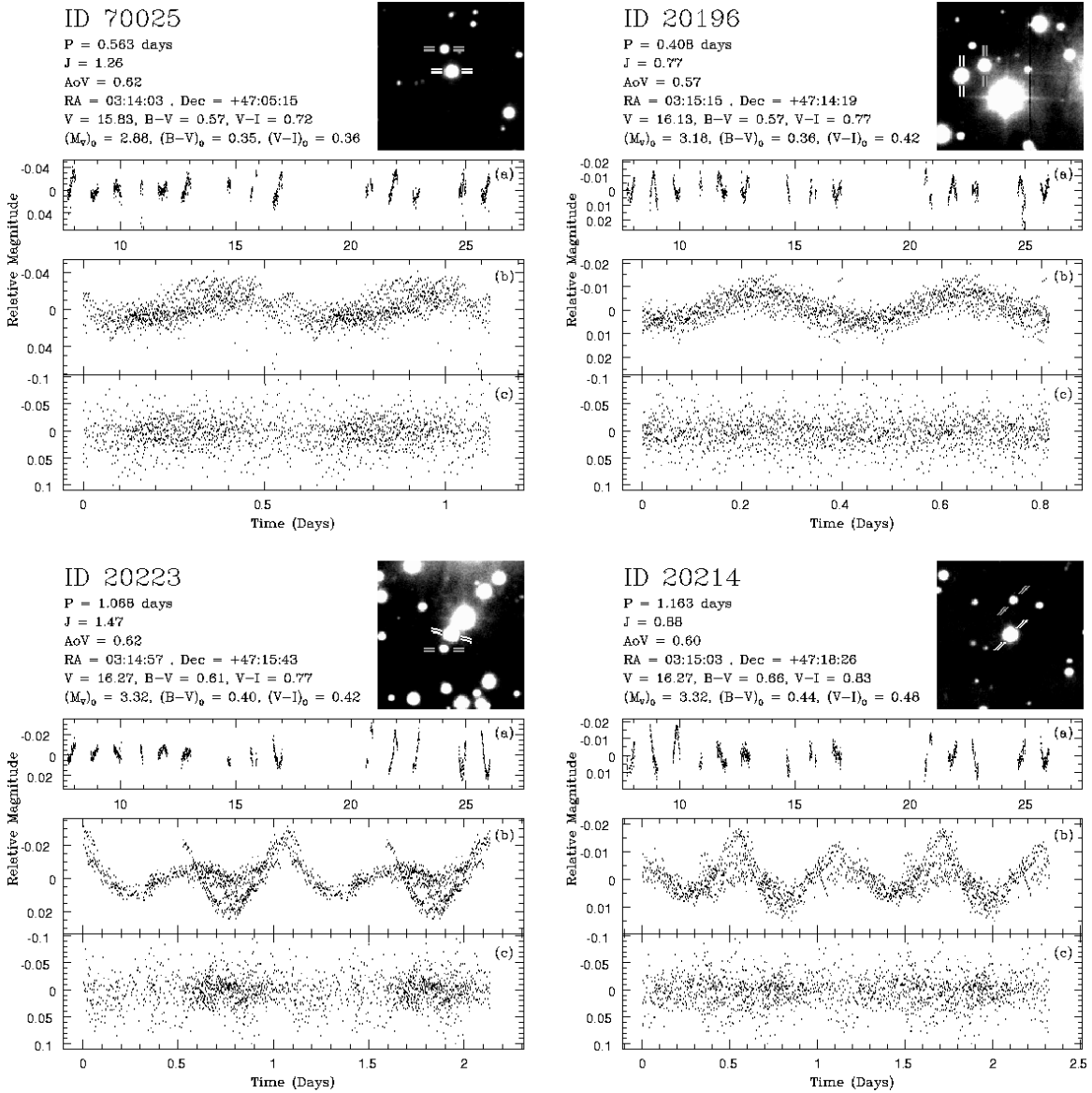}
\caption{Light curves and finder charts for variables that are potential cluster members.  In the top right is the 
finder chart, with the variable marked with the white bars and the nearest detected object 
marked with gray bars.  Panel (a) shows the full unphased light curve for the 
variable.  Panel (b) shows the variable's light curve phased to the nearest 
period, with two periods plotted.  Panel (c) shows the light curve of the nearest 
object phased to the period of the variable.}
\label{fig:p0}
\end{figure}

We show the light curves and photometric properties of the potential
cluster members meeting both selection cuts in Figures \ref{fig:p0}
though \ref{fig:p4}.  The plots for the field variables are in Figures
\ref{fig:p5} though \ref{fig:p10}.  For each variable we show the
unphased lightcurve in panel (a) for the full 19 days of the survey.
In panel (b) we show the phased lightcurve of the variable, with two
periods plotted.  In the upper right corner is a finding chart showing
the variable marked with white bars.  The nearest detected object is
marked with gray bars, and the light curve for the nearest object is
shown in panel (c), phased to the period of the variable.  We show the
phased lightcurve for the nearest source because some of the
millimagnitude amplitude variables are visually blended.  Having an
object nearby with similar blending complications that does not reveal
variability confirms the intrinsic nature of the low amplitude
variable.  For completeness, we also show light curves in Figure
\ref{fig:low_vars} for two objects that do not survive the second tier
cuts, but visually demonstrate intrinsic variability.

It should be noted that the DoPhot photometry detection software
cannot successfully identify every object in the images.  Objects that
are too heavily blended, elongated, faint, saturated, etc., do not
make it through the photometry pipeline.  Therefore, the nearest
object identified by the software may not be the nearest object to the
variable.  In most cases, however, the object identified as nearest is
still close enough to the variable for the light curve comparison to
be useful.

We plot the properties of the potential cluster variables in Table
\ref{tab:mvars}, and the field variables in Table \ref{tab:fvars},
with the two visually selected stars listed separately.  For a
variable to belong to the cluster it must have $BVI$ apparent
magnitudes consistent with the cluster main sequence.  For potential 
cluster members, we give
estimates for the stellar properties based on the CMD (see
\S\ref{sec:membership}), and we classify their variable types (see
\S\ref{sec:vars}) if possible.  Both tables give magnitudes and
positional data for all the variable stars.

\begin{figure}
\epsscale{1.0}
\plotone{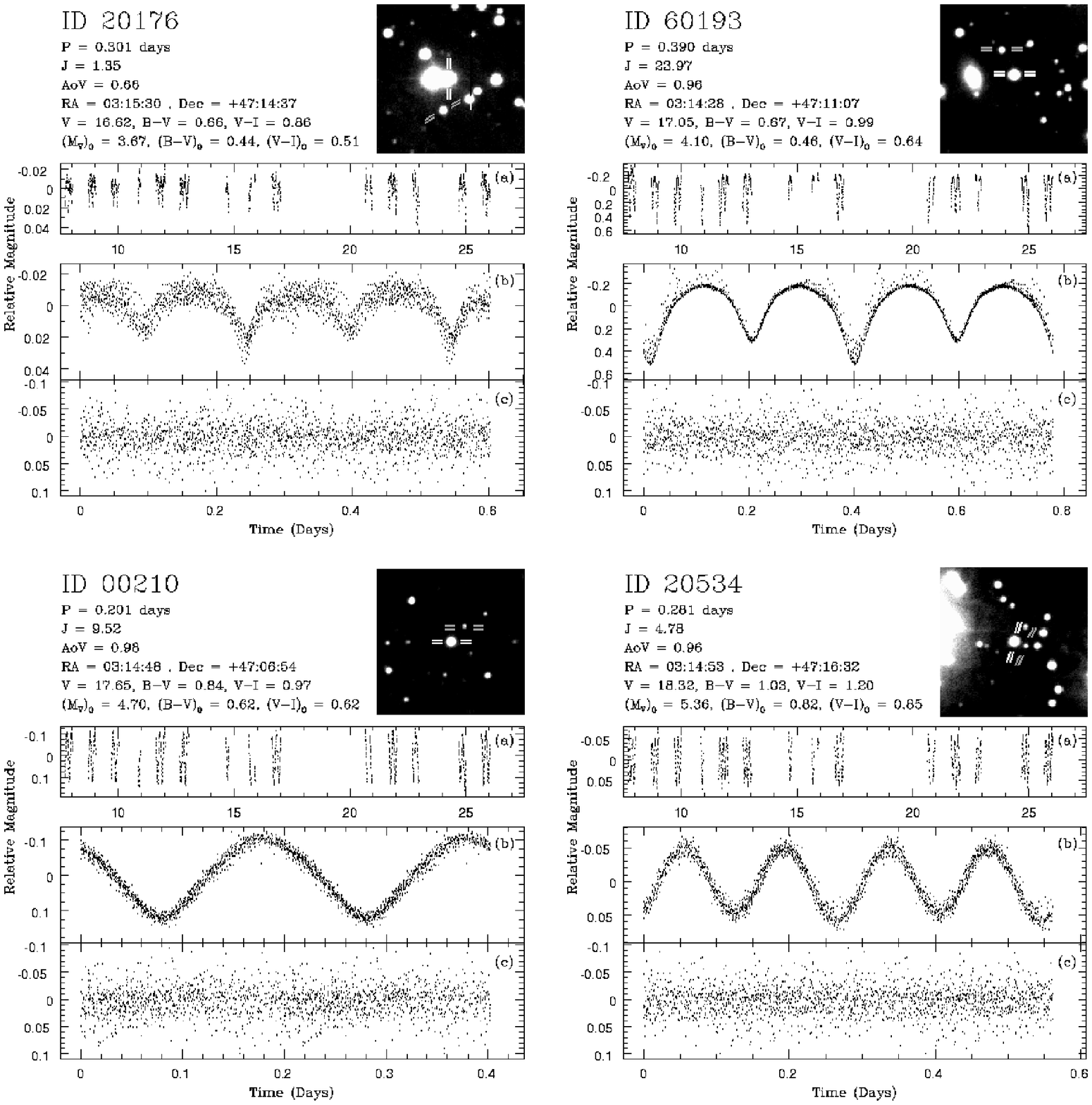}
\caption{Light curves and finder charts for potential cluster variables.  See caption of Figure \ref{fig:p0} for description.}
\label{fig:p1}
\end{figure}

\begin{figure}
\epsscale{1.0}
\plotone{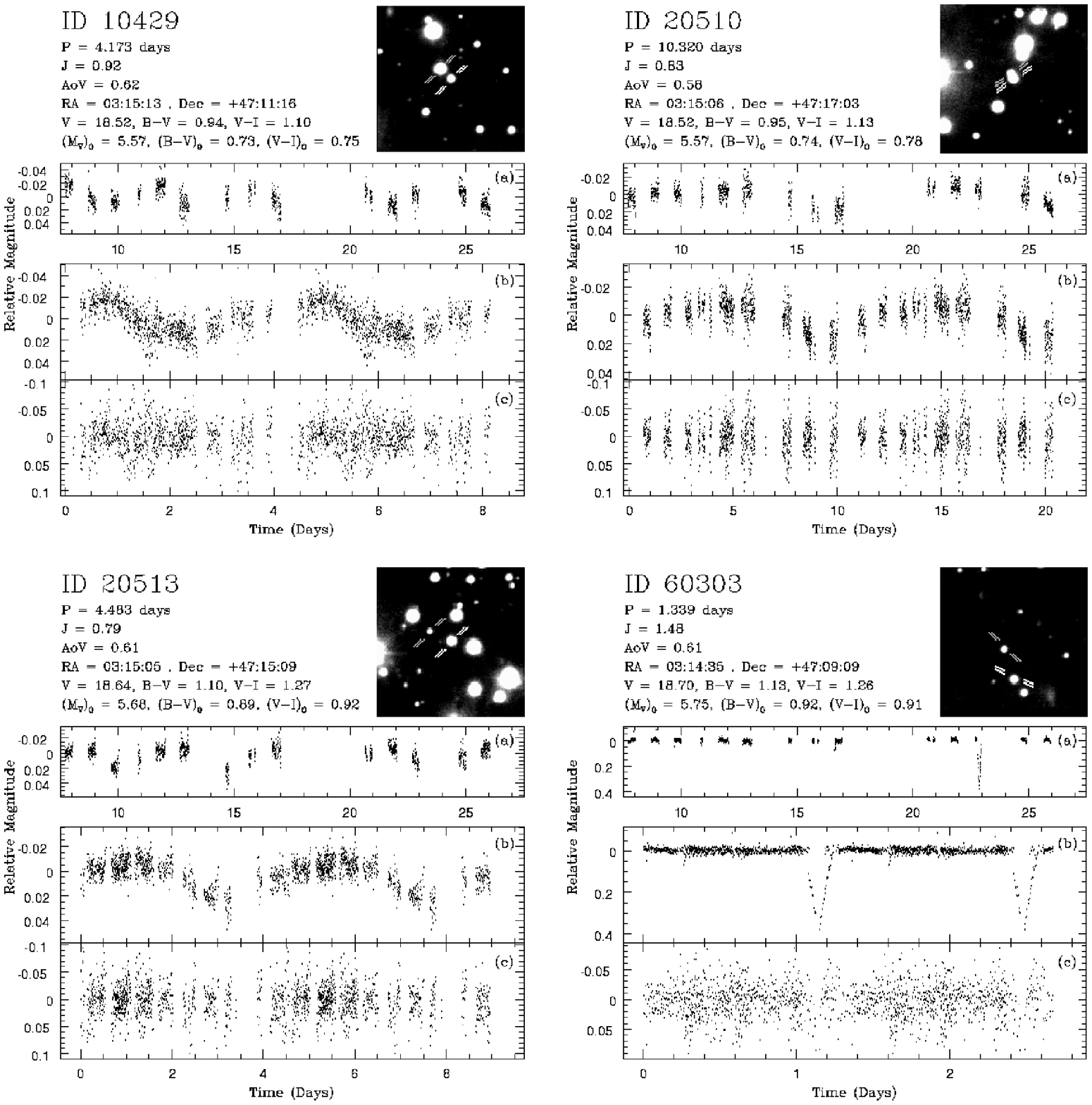}
\caption{Light curves and finder charts for potential cluster variables.  See caption of Figure \ref{fig:p0} for description.}
\label{fig:p2}
\end{figure}

\begin{figure}
\epsscale{1.0}
\plotone{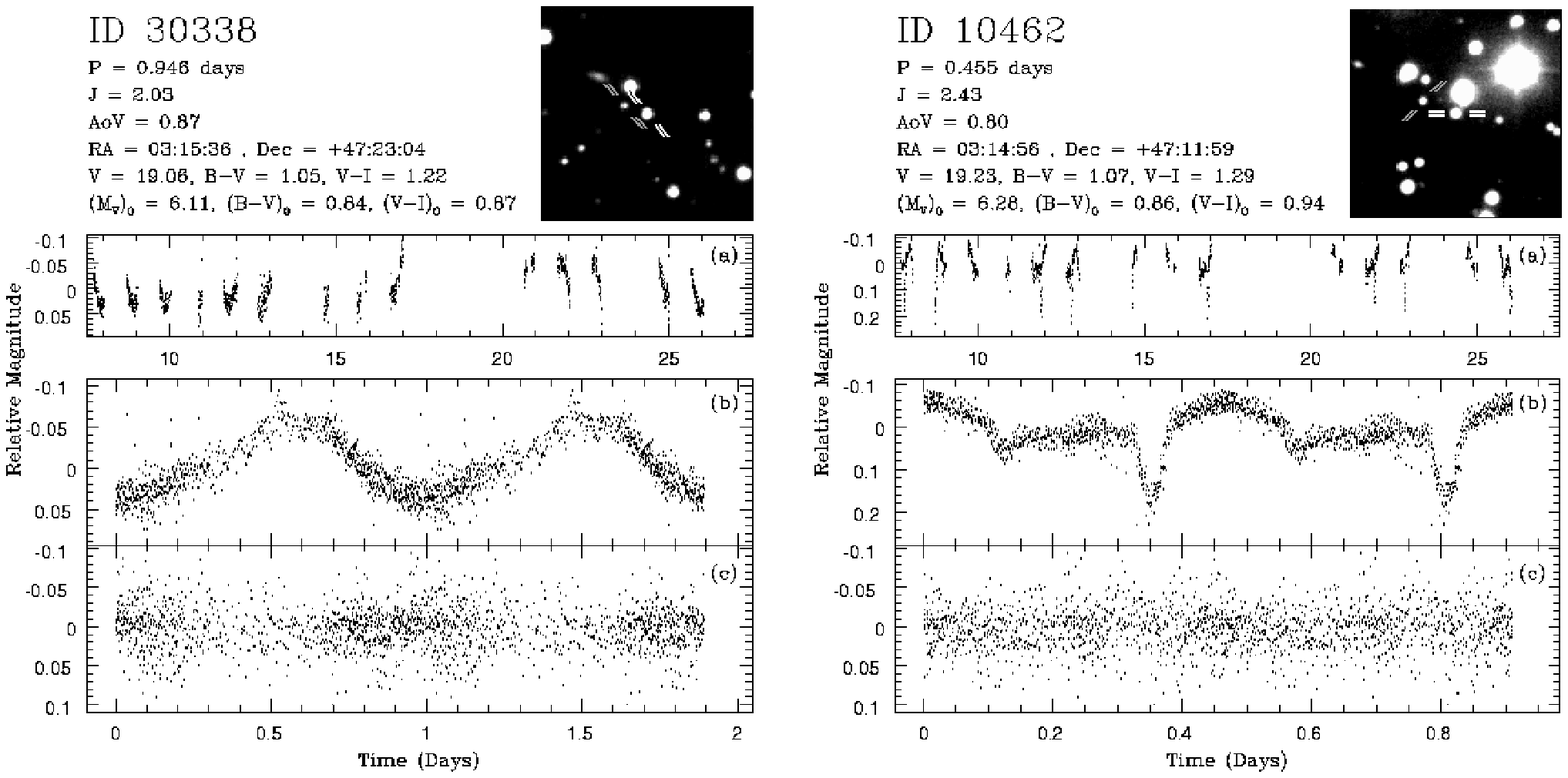}
\caption{Light curves and finder charts for potential cluster variables.  See caption of Figure \ref{fig:p0} for description.}
\label{fig:p3}
\end{figure}

\begin{figure}
\epsscale{1.0}
\plotone{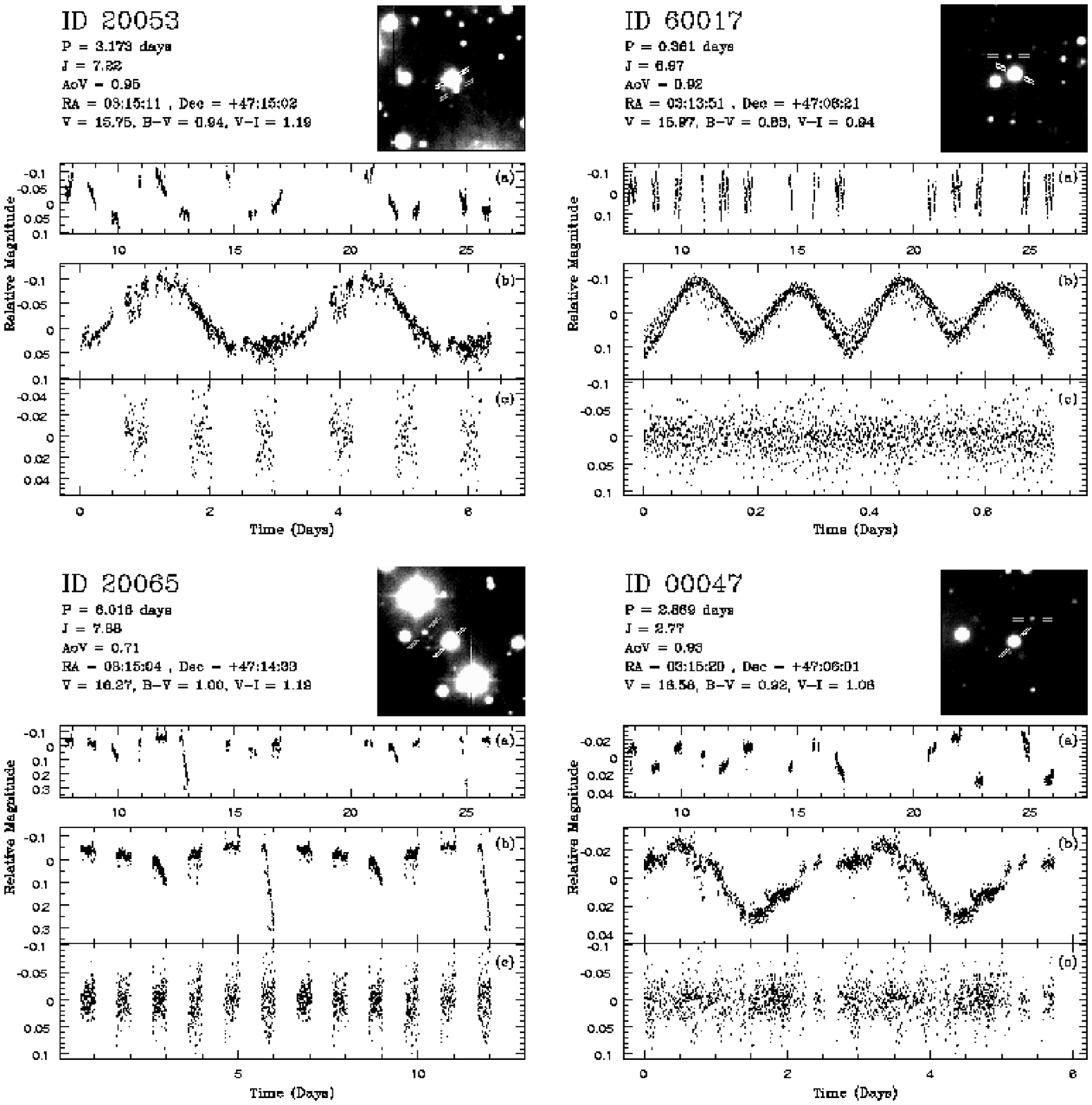}
\caption{Light curves and finder charts for field star variables.  See caption of Figure \ref{fig:p0} for description.}
\label{fig:p4}
\end{figure}

\begin{figure}
\epsscale{1.0}
\plotone{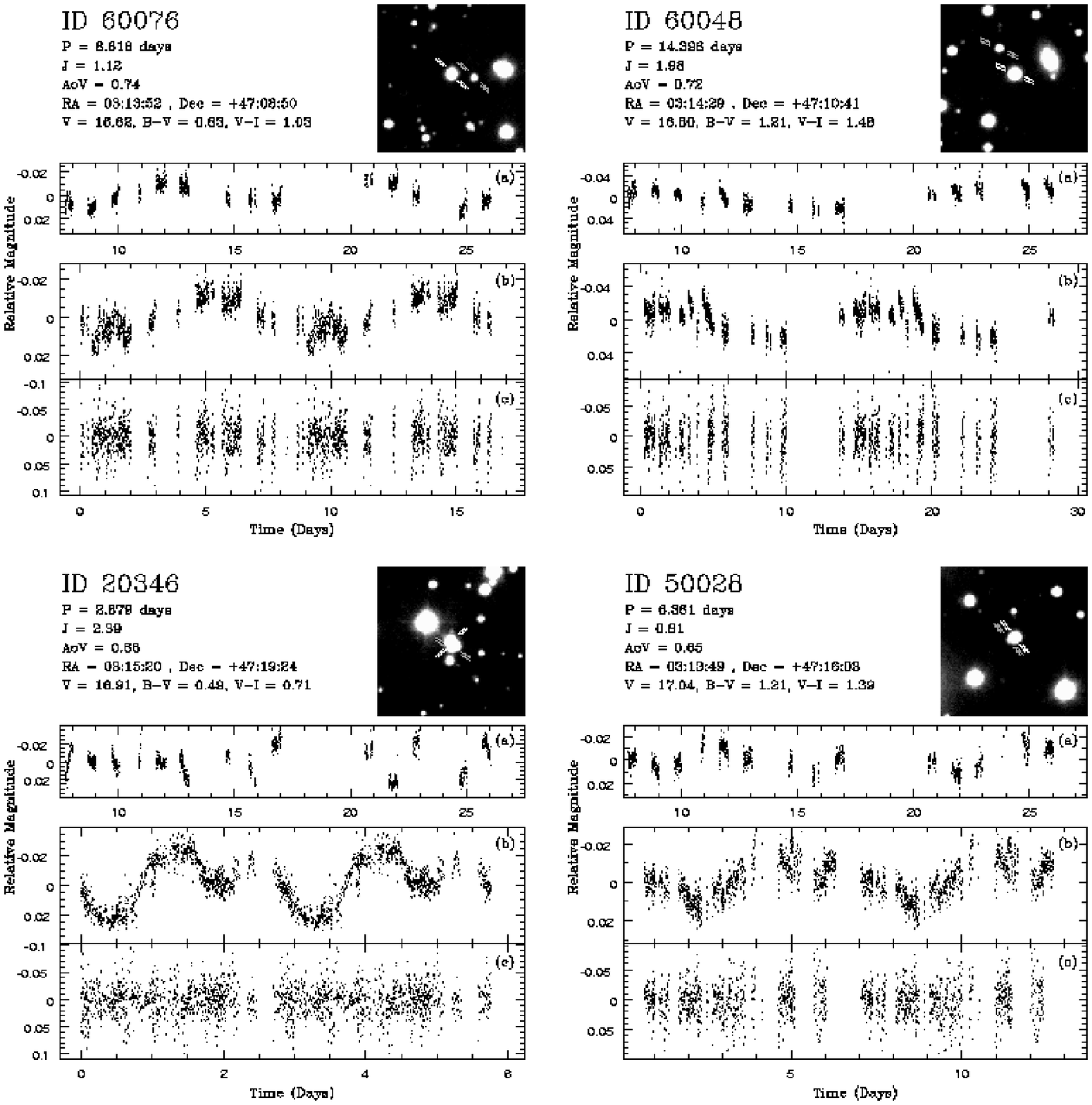}
\caption{Light curves and finder charts for field star variables.  See caption of Figure \ref{fig:p0} for description.}
\label{fig:p5}
\end{figure}

\begin{figure}
\epsscale{1.0}
\plotone{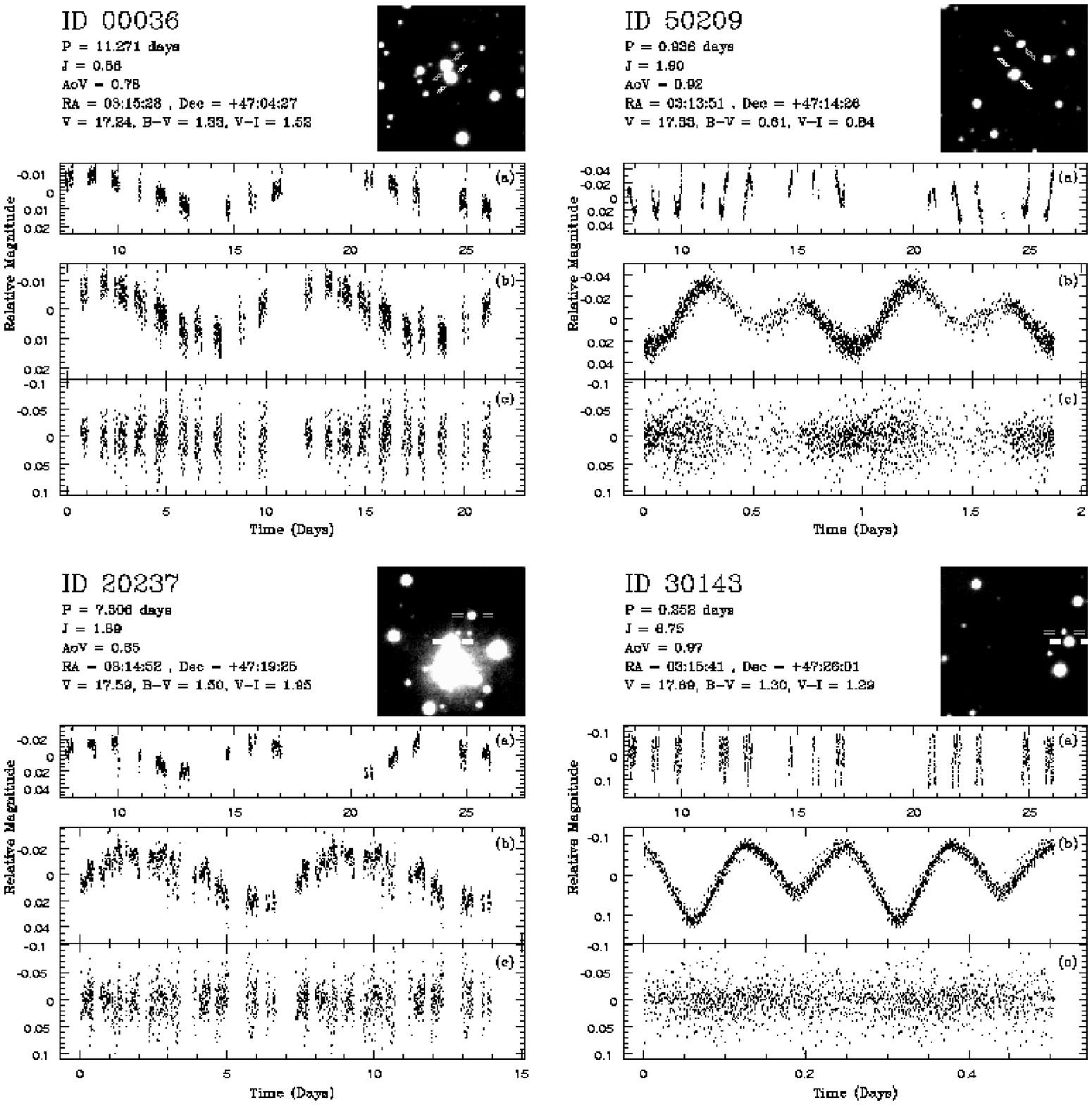}
\caption{Light curves and finder charts for field star variables.  See caption of Figure \ref{fig:p0} for description.}
\label{fig:p6}
\end{figure}

\begin{figure}
\epsscale{1.0}
\plotone{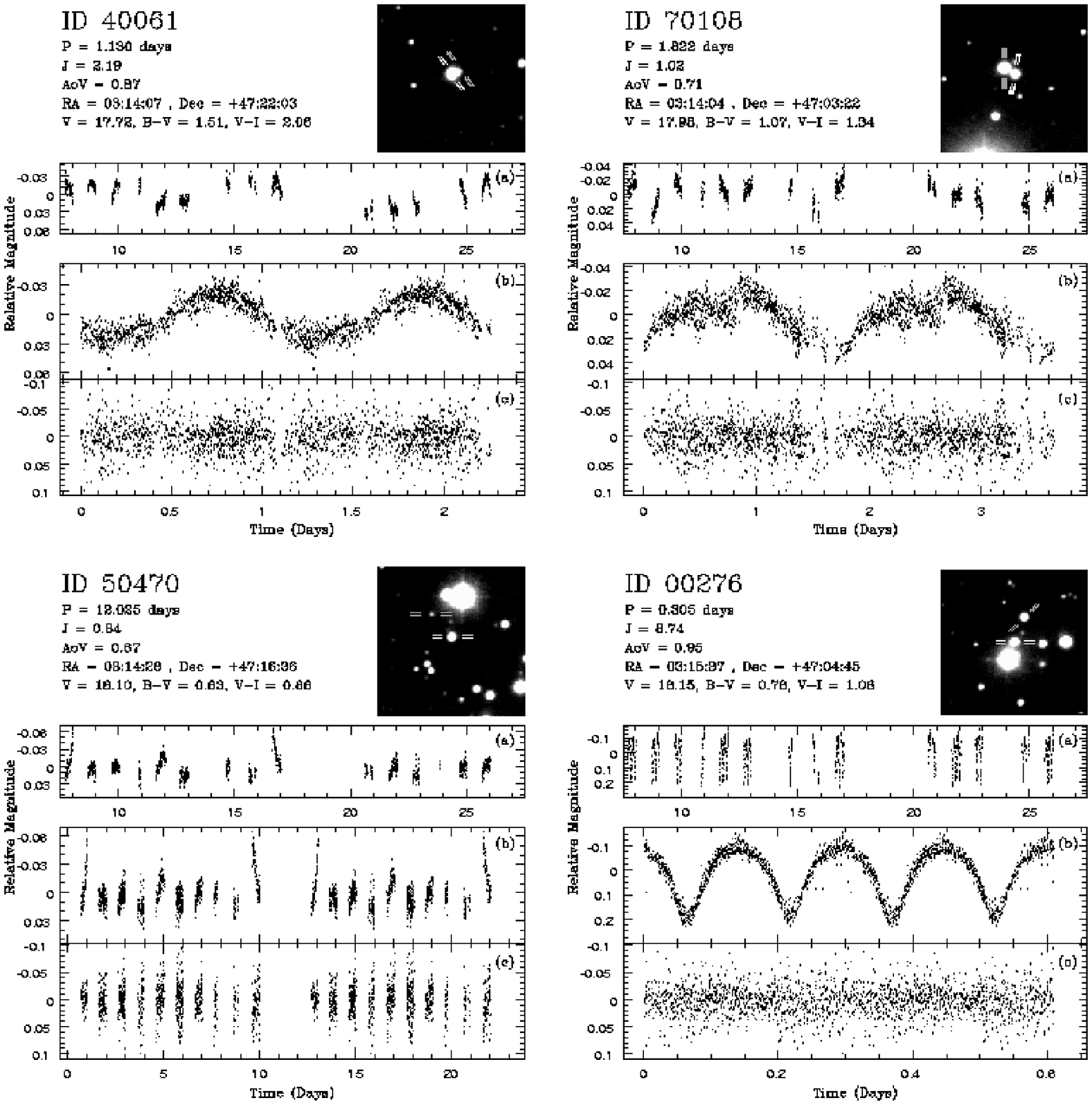}
\caption{Light curves and finder charts for field star variables.  See caption of Figure \ref{fig:p0} for description.}
\label{fig:p7}
\end{figure}

\begin{figure}
\epsscale{1.0}
\plotone{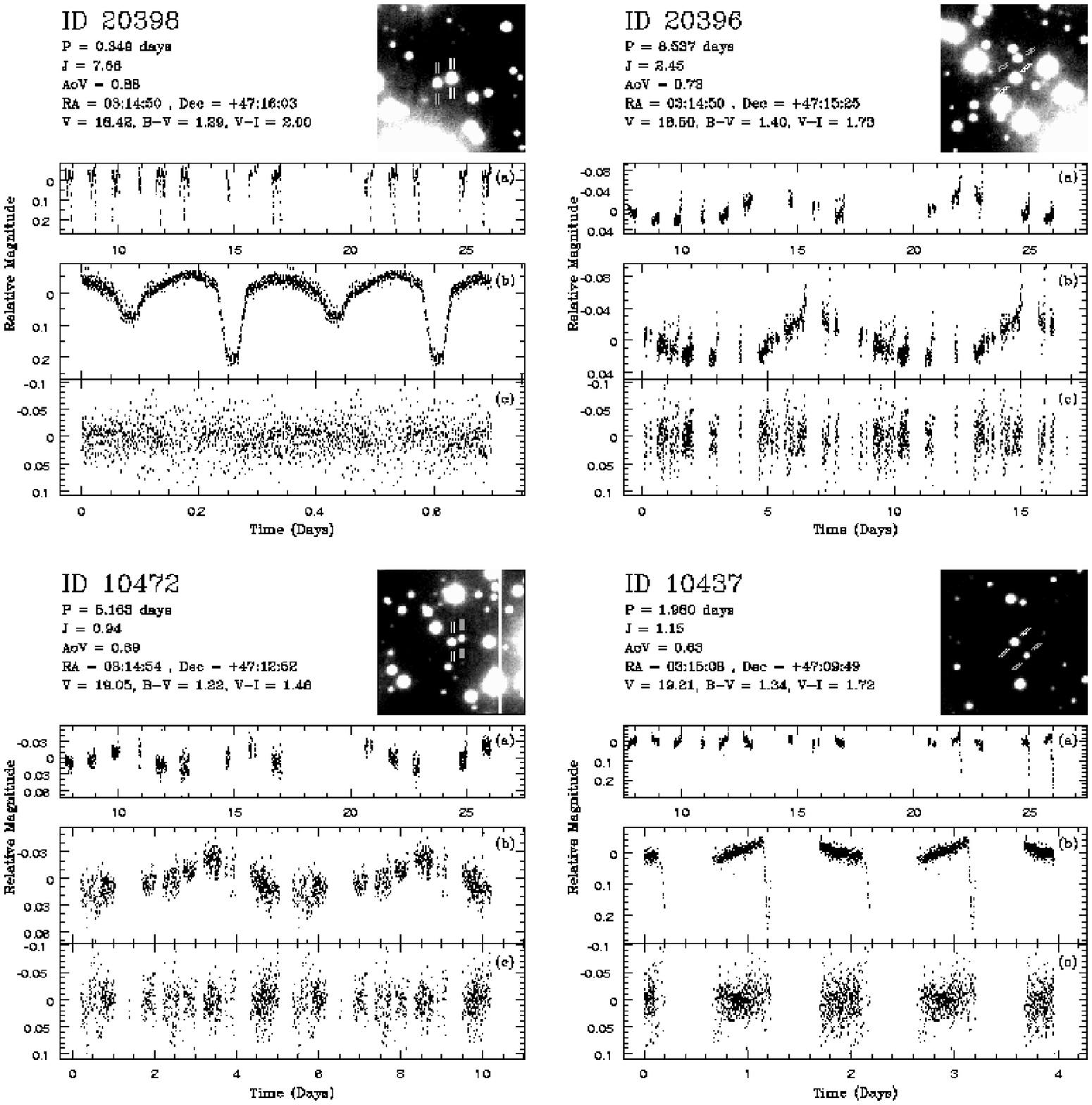}
\caption{Light curves and finder charts for field star variables.  See caption of Figure \ref{fig:p0} for description.}
\label{fig:p8}
\end{figure}

\begin{figure}
\epsscale{1.0}
\plotone{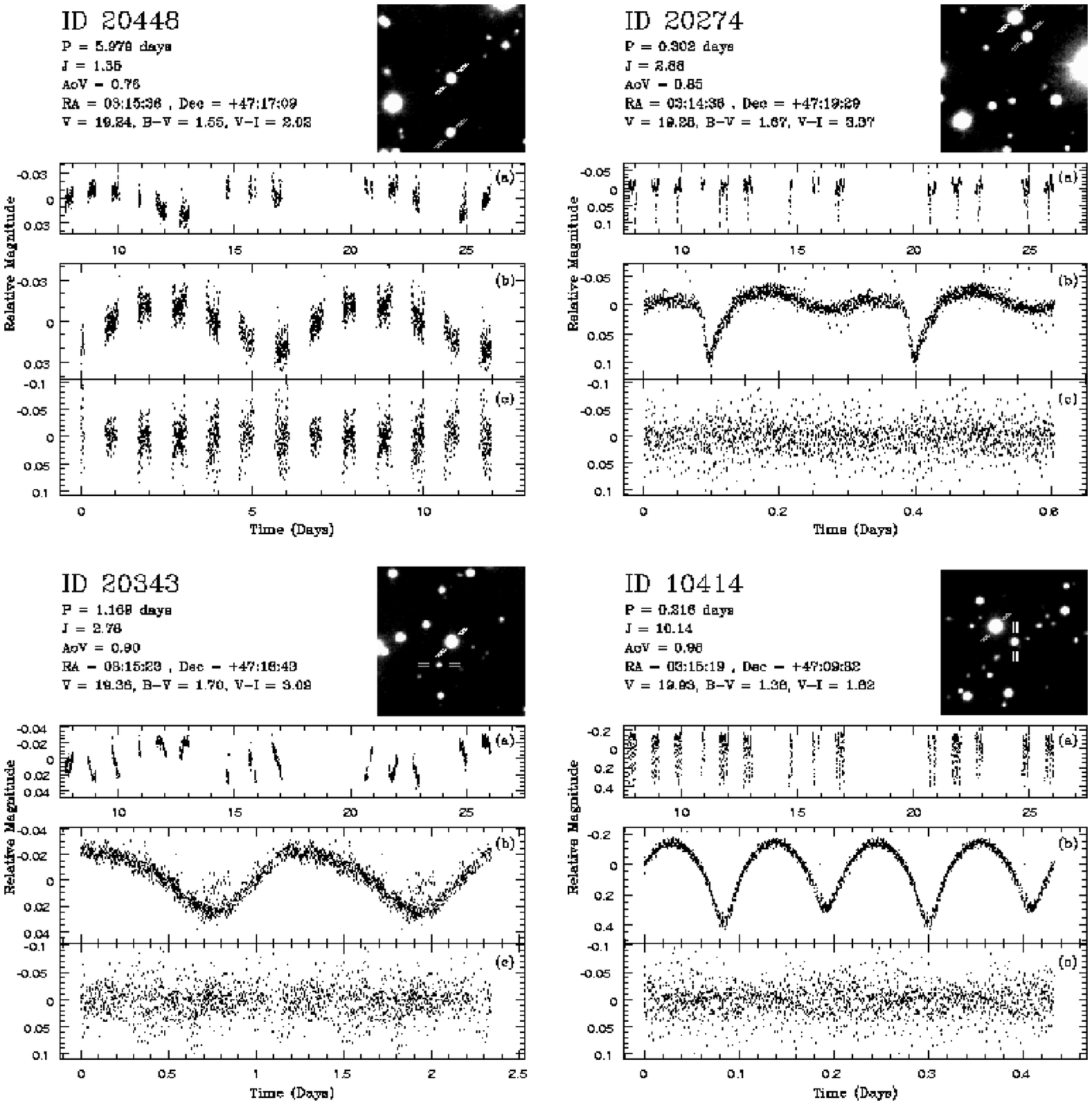}
\caption{Light curves and finder charts for field star variables.  See caption of Figure \ref{fig:p0} for description.}
\label{fig:p9}
\end{figure}

\begin{figure}
\epsscale{1.0}
\plotone{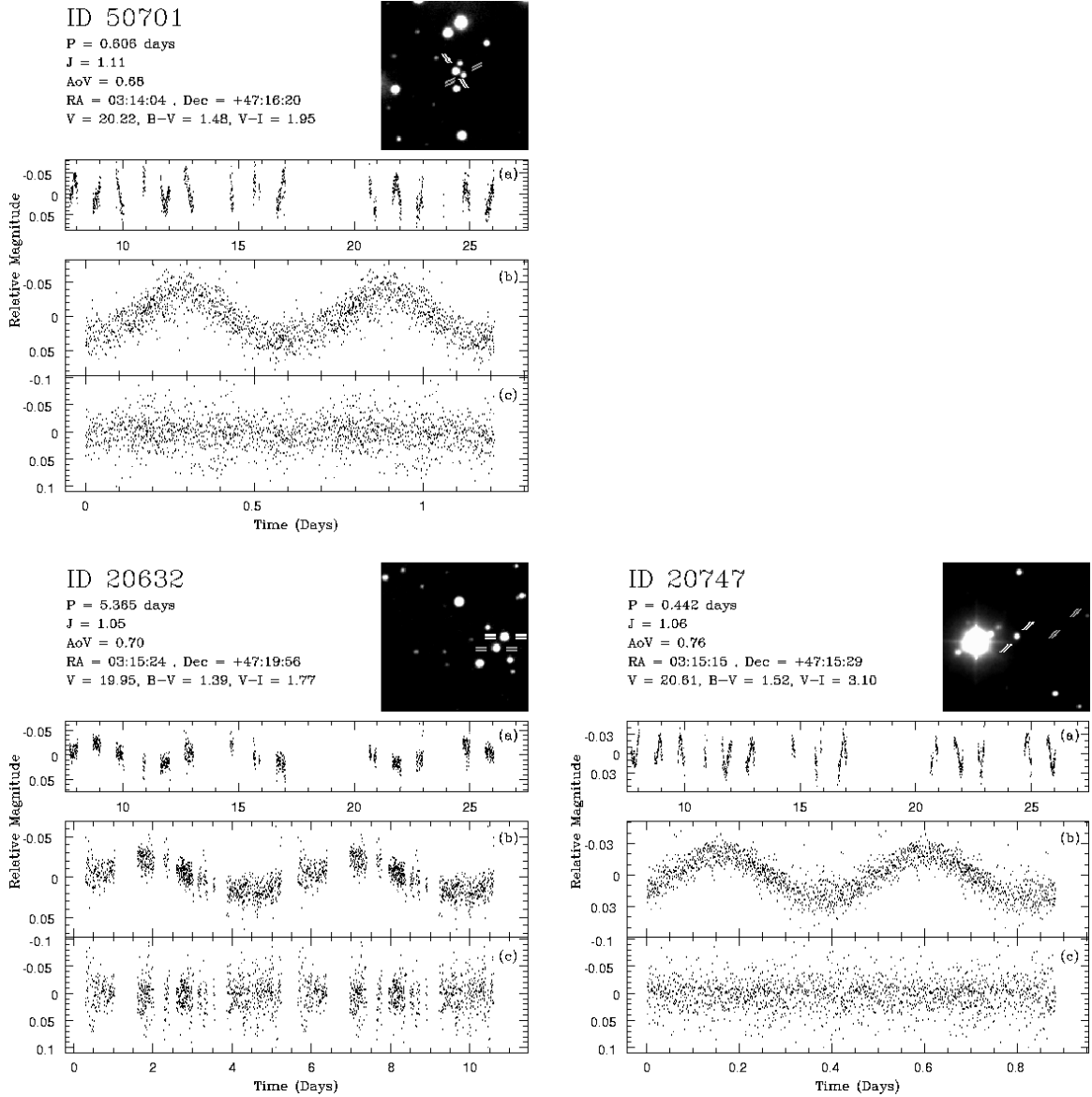}
\caption{Light curves and finder charts for field star variables.  See caption of Figure \ref{fig:p0} for description.  The brightness scaling for the finder chart for object 20747 is different from the other finder charts, with objects appearing dimmer, so that the bright star in the left of the chart does not swamp the variable.  For that reason, the nearest identified star is not visible in this scaling.}
\label{fig:p10}
\end{figure}

\begin{figure}
\epsscale{1.0}
\plotone{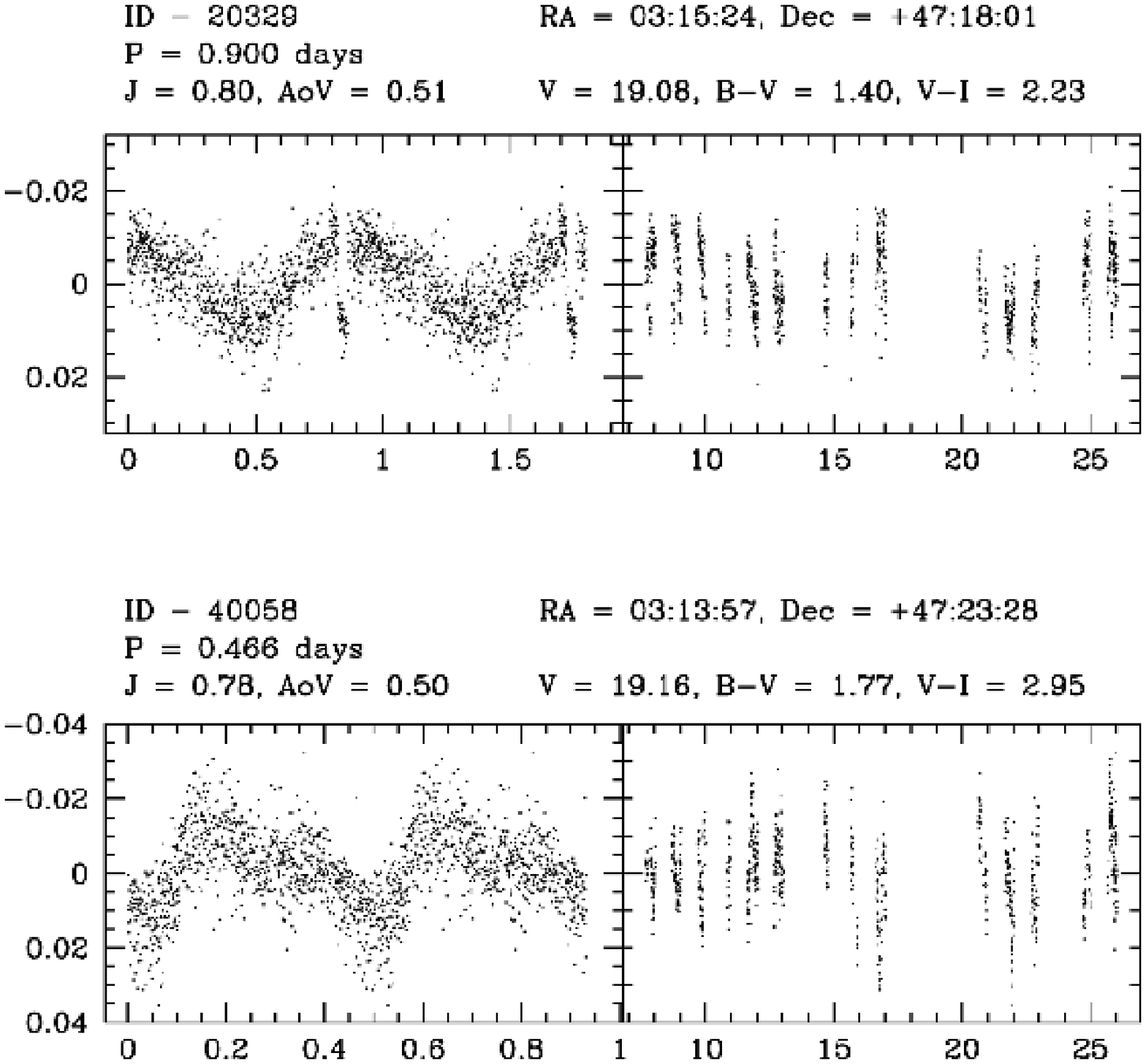}
\caption{Light curves for two objects that did not survive the variability cuts, but show coherent 
periodic variability upon visual inspection.  Phased lightcurves showing two full periods are shown 
on the left, and the full unphased light curves are shown on the right.}
\label{fig:low_vars}
\end{figure}

\begin{deluxetable}{ccccccccccccc}
\tabletypesize{\footnotesize}
\tablecaption{Properties of Variable Potential Cluster Members\label{tab:mvars}}
\tablewidth{0pt}
\tablehead{  \colhead{STEPSS} & \colhead{$(M_V)_0$}& \colhead{$(B-V)_0$} & \colhead{$(V-I)_0$} & \colhead{RA} & \colhead{Dec} & \colhead{Variable} & \colhead{$M$} & \colhead{$R$} & \colhead{($T_{eff}$)} & \colhead{$J$} & \colhead{$AoV$} & \colhead{Period} \\ \colhead{ID \#} & \colhead{\tablenotemark{a} } & \colhead{\tablenotemark{b}} & \colhead{\tablenotemark{b}} & \colhead{(2000.0)} & \colhead{(2000.0)} & \colhead{Type} & \colhead{($M_\odot$)} & \colhead{($R_\odot$)} & \colhead{(K)} & \colhead{} & \colhead{} & \colhead{(days)} }
\startdata
\hline
70025 & 2.88 & 0.35 & 0.36 & 03 14 03 & +47 05 15 & $\gamma$ Dor & 1.47 & 1.56 & 7000 & 1.262  & 0.619 & 0.563  \\
20196 & 3.18 & 0.36 & 0.42 & 03 15 15 & +47 14 19 & $\gamma$ Dor & 1.40 & 1.45 & 6800 & 0.774  & 0.566 & 0.408  \\
20223 & 3.32 & 0.40 & 0.42 & 03 14 57 & +47 15 43 & $\gamma$ Dor & 1.36 & 1.40 & 6700 & 1.470  & 0.619 & 1.068  \\
20214 & 3.32 & 0.44 & 0.48 & 03 15 03 & +47 18 26 & $\gamma$ Dor & 1.36 & 1.40 & 6700 & 0.877  & 0.604 & 1.163  \\
20176 & 3.67 & 0.44 & 0.51 & 03 15 30 & +47 14 37 & EW           & 1.28 & 1.28 & 6500 & 1.352  & 0.656 & 0.301  \\
60193 & 4.10 & 0.46 & 0.64 & 03 14 28 & +47 11 07 & EW           & 1.19 & 1.16 & 6300 & 23.966 & 0.962 & 0.390  \\
00210 & 4.70 & 0.62 & 0.62 & 03 14 48 & +47 06 54 & --           & 1.07 & 0.98 & 5900 & 9.515  & 0.977 & 0.201  \\
20534 & 5.36 & 0.82 & 0.85 & 03 14 53 & +47 16 32 & EW           & 0.96 & 0.85 & 5600 & 4.777  & 0.960 & 0.281  \\
10429 & 5.57 & 0.73 & 0.75 & 03 15 13 & +47 11 16 & --           & 0.93 & 0.82 & 5400 & 0.921  & 0.623 & 4.173  \\
20510 & 5.57 & 0.74 & 0.78 & 03 15 06 & +47 17 03 & --           & 0.93 & 0.82 & 5400 & 0.827  & 0.584 & 10.32  \\
20513 & 5.69 & 0.89 & 0.92 & 03 15 05 & +47 15 09 & --           & 0.91 & 0.80 & 5400 & 0.794  & 0.611 & 4.483  \\
60303 & 5.76 & 0.92 & 0.91 & 03 14 35 & +47 09 09 & EA           & 0.89 & 0.79 & 5300 & 1.478  & 0.609 & 1.339  \\
30338 & 6.11 & 0.84 & 0.87 & 03 15 36 & +47 23 04 & --           & 0.85 & 0.75 & 5100 & 2.034  & 0.873 & 0.946  \\
10462 & 6.28 & 0.86 & 0.94 & 03 14 56 & +47 11 59 & EA           & 0.83 & 0.73 & 5100 & 2.434  & 0.799 & 0.455  \\
\hline
\enddata
\tablenotetext{a}{$(M_V)_0$ determined from observed $V$ magnitude using $A_V = 0.68 \pm 0.09$ and distance modulus of $12.27 \pm 0.12$ from \citet{burke04}.  Combined instrumental and systematic errors in $(M_V)_0$ are about 0.15 mag.}
\tablenotetext{b}{Dereddened values for $(B-V)_0$ and $(V-I)_0$ determined using $A_V = 0.68$, $A_I = 0.33$, with $R_V = 3.2$, from \citet{burke04}.  Combined instrumental and systematic errors in the colors are about 0.1 mag in $(B-V)_0$ and 0.15 in $(V-I)_0$.}
			    
\end{deluxetable}

\begin{deluxetable}{ccccccccc}
\tabletypesize{\footnotesize}
\tablecaption{Properties of Variable Field Stars\label{tab:fvars}}
\tablewidth{0pt}
\tablehead{ \colhead{STEPSS ID \#} & \colhead{$V$}\tablenotemark{a} & \colhead{$B-V$}\tablenotemark{a} & \colhead{$V-I$}\tablenotemark{b} & \colhead{RA} & \colhead{Dec} & \colhead{$J$} & \colhead{$AoV$} & \colhead{Period} \\ \colhead{} & \colhead{} & \colhead{} & \colhead{} & \colhead{(2000.0)} & \colhead{(2000.0)} & \colhead{} & \colhead{} & \colhead{(days)} }
\startdata
\hline
20053  & 15.75 & 0.95 & 1.18 & 03 15 11 & +47 15 02 & 7.223  & 0.945 & 3.173 \\
60017  & 15.98 & 0.82 & 0.95 & 03 13 51 & +47 08 21 & 6.974  & 0.924 & 0.361 \\
20065  & 16.28 & 0.99 & 1.20 & 03 15 04 & +47 14 33 & 7.879  & 0.707 & 6.016 \\
00047  & 16.59 & 0.91 & 1.06 & 03 15 20 & +47 06 01 & 2.770  & 0.931 & 2.869 \\
60076  & 16.62 & 0.83 & 1.03 & 03 13 52 & +47 08 50 & 1.118  & 0.736 & 8.618 \\
60048  & 16.80 & 1.20 & 1.48 & 03 14 29 & +47 10 41 & 1.979  & 0.721 & 14.40 \\
20346  & 16.91 & 0.49 & 0.71 & 03 15 20 & +47 19 24 & 2.387  & 0.876 & 2.879 \\
50028  & 17.04 & 1.21 & 1.39 & 03 13 49 & +47 16 03 & 0.808  & 0.652 & 6.361 \\
00036  & 17.24 & 1.33 & 1.52 & 03 15 28 & +47 04 27 & 0.856  & 0.784 & 11.27 \\
50209  & 17.33 & 0.61 & 0.84 & 03 13 51 & +47 14 26 & 1.901  & 0.916 & 0.936 \\
20237  & 17.59 & 1.50 & 1.95 & 03 14 52 & +47 19 25 & 1.887  & 0.846 & 7.306 \\
30143  & 17.69 & 1.30 & 1.29 & 03 15 41 & +47 26 01 & 6.753  & 0.974 & 0.252 \\
40061  & 17.72 & 1.51 & 2.06 & 03 14 07 & +47 22 03 & 2.191  & 0.868 & 1.130 \\
70108  & 17.99 & 1.07 & 1.35 & 03 14 04 & +47 03 22 & 1.019  & 0.711 & 1.822 \\
50470  & 18.10 & 0.62 & 0.86 & 03 14 28 & +47 16 36 & 0.843  & 0.666 & 12.03 \\
00276  & 18.15 & 0.78 & 1.06 & 03 15 37 & +47 04 45 & 8.737  & 0.953 & 0.304 \\
20398  & 18.42 & 1.29 & 2.00 & 03 14 50 & +47 16 03 & 7.660  & 0.885 & 0.349 \\
20396  & 18.50 & 1.40 & 1.73 & 03 14 50 & +47 15 25 & 2.449  & 0.731 & 8.537 \\
10472  & 19.05 & 1.23 & 1.45 & 03 14 54 & +47 12 52 & 0.939  & 0.688 & 5.163 \\
10437  & 19.21 & 1.34 & 1.72 & 03 15 08 & +47 09 49 & 1.145  & 0.633 & 1.980 \\
20448  & 19.24 & 1.56 & 2.02 & 03 15 36 & +47 17 09 & 1.349  & 0.764 & 5.979 \\
20274  & 19.28 & 1.66 & 3.37 & 03 14 36 & +47 19 29 & 2.880  & 0.854 & 0.302 \\
20343  & 19.36 & 1.70 & 3.09 & 03 15 23 & +47 18 43 & 2.758  & 0.899 & 1.169 \\
10414  & 19.93 & 1.38 & 1.82 & 03 15 19 & +47 09 32 & 10.143 & 0.980 & 0.216 \\
20632  & 19.95 & 1.39 & 1.78 & 03 15 24 & +47 19 56 & 1.051  & 0.700 & 5.365 \\
50701  & 20.22 & 1.48 & 1.95 & 03 14 04 & +47 16 20 & 1.111  & 0.682 & 0.606 \\
20747  & 20.61 & 1.52 & 3.11 & 03 15 15 & +47 15 29 & 1.056  & 0.756 & 0.442 \\ \hline
20329  & 19.08 & 1.40 & 2.23 & 03 15 24 & +47 18 01 & 0.796  & 0.511 & 0.900 \\
40058  & 19.16 & 1.77 & 2.95 & 03 13 57 & +47 23 28 & 0.782  & 0.503 & 0.931 \\
\enddata
\tablenotetext{a}{Combined instrumental and systematic errors in absolute $B$ and $V$ magnitude are 0.05 mag.}
\tablenotetext{b}{Combined instrumental and systematic errors in absolute $I$ magnitude are 0.1 mag.}
			    
\end{deluxetable}

\subsection{Matching To 2MASS} \label{sec:2mass} 

Most of the selected variables have
matching counterparts in the 2MASS database \citep{sk97}, which we accessed through 
the VizieR web portal\footnote{http://vizier.u-strasbg.fr/viz-bin/VizieR}.  For variables with a
2MASS counterpart within 1$\arcsec$ radius, we list the 2MASS ID numbers, the
2MASS magnitudes in the $J$, $H$, and $K$ bands, and the distance
between the variable positions we measure and the reported positions
of the 2MASS sources.  Three of the faintest stars, two cluster members and one 
field star, were not matched to any 2MASS object.  A search of the literature revealed 
no known variable counterpart to any of the variable stars identified in this 
analysis.  We were unable to find any published cases of variable stars in the 
observed field within the magnitude ranges we searched.  We therefore conclude that 
all of the variables identified in this paper are previously unknown.

\begin{deluxetable}{cccccc}
\tabletypesize{\footnotesize}
\tablecaption{Variable Matches to 2MASS \label{tab:2mass}}
\tablewidth{0pt}
\tablehead{  \colhead{STEPSS ID} & \colhead{2MASS ID} & \colhead{2MASS Offset} & \colhead{$J$} & \colhead{$H$} & \colhead{$K$} \\
\colhead{} & \colhead{} & \colhead{(arcsec)} & \colhead{} & \colhead{} & \colhead{}
}
\startdata
 \multicolumn{6}{c}{Cluster Members}\\
\hline
00210 & 03144786+4706544 & 0.098 & 16.116 $\pm$ 0.094 & 15.670 $\pm$ 0.120 & 15.881 $\pm$ 0.065\\
20534 & 03145325+4716321 & 0.579 & 16.270 $\pm$ 0.107 & 15.683 $\pm$ 0.124 & 15.512 $\pm$ 0.172\\
20176 & 03152956+4714374 & 0.066 & 15.240 $\pm$ 0.044 & 14.901 $\pm$ 0.059 & 14.677 $\pm$ 0.069\\
60193 & 03142811+4711067 & 0.079 & 15.757 $\pm$ 0.072 & 15.245 $\pm$ 0.091 & 15.348 $\pm$ 0.158\\
20196 & 03151490+4714185 & 0.043 & 14.904 $\pm$ 0.030 & 14.670 $\pm$ 0.051 & 14.555 $\pm$ 0.060\\
10429 & 03151250+4711157 & 0.232 & 16.635 $\pm$ 0.149 & 15.760 $\pm$ 0.068 & 14.873 $\pm$ 0.101\\
10462 & ---   	        &       &    	  	     &       	          &                   \\
70025 & 03140339+4705144 & 0.115 & 14.747 $\pm$ 0.040 & 14.509 $\pm$ 0.051 & 14.483 $\pm$ 0.069\\
30338 & ---   		&       &    	  	     &       	          &                   \\
20223 & 03145747+4715430 & 0.063 & 14.922 $\pm$ 0.043 & 14.733 $\pm$ 0.065 & 14.531 $\pm$ 0.080\\
20214 & 03150277+4718258 & 0.063 & 14.939 $\pm$ 0.033 & 14.621 $\pm$ 0.048 & 14.598 $\pm$ 0.077\\
60303 & 03143504+4709095 & 0.363 & 16.608 $\pm$ 0.150 & 16.140 $\pm$ 0.180 & 15.785 $\pm$ 0.216\\
20513 & 03150464+4715090 & 0.356 & 16.438 $\pm$ 0.117 & 15.938 $\pm$ 0.153 & 16.060 $\pm$ 0.030\\
20510 & 03150636+4717030 & 0.572 & 16.396 $\pm$ 0.116 & 16.033 $\pm$ 0.168 & 15.821 $\pm$ 0.230\\
\hline
 \multicolumn{6}{c}{Field Stars}\\
\hline
10414 & 03151946+4709324 & 0.483 & 17.050 $\pm$ 0.151 & 16.311 $\pm$ 0.173 & 16.243 $\pm$ 0.283\\
30143 & 03154104+4726009 & 0.109 & 15.495 $\pm$ 0.048 & 14.874 $\pm$ 0.064 & 14.804 $\pm$ 0.095\\
20274 & 03143588+4719290 & 0.056 & 14.097 $\pm$ 0.025 & 13.486 $\pm$ 0.027 & 13.140 $\pm$ 0.030\\
00276 & 03153653+4704457 & 0.483 & 16.386 $\pm$ 0.096 & 15.946 $\pm$ 0.156 & 15.708 $\pm$ 0.198\\
20398 & 03144954+4716030 & 0.124 & 15.156 $\pm$ 0.049 & 14.508 $\pm$ 0.053 & 14.259 $\pm$ 0.061\\
60017 & 03135090+4708209 & 0.079 & 14.449 $\pm$ 0.034 & 14.054 $\pm$ 0.045 & 13.983 $\pm$ 0.049\\
20747 & 03151453+4715287 & 0.051 & 15.779 $\pm$ 0.074 & 15.178 $\pm$ 0.079 & 14.925 $\pm$ 0.088\\
50701 & ---		&       &    	  	     &       	          &                   \\
50209 & 03135130+4714258 & 0.123 & 15.984 $\pm$ 0.078 & 15.727 $\pm$ 0.130 & 15.145 $\pm$ 0.135\\
40061 & 03140666+4722034 & 0.094 & 14.462 $\pm$ 0.031 & 13.657 $\pm$ 0.033 & 13.494 $\pm$ 0.040\\
20343 & 03152256+4718432 & 0.027 & 14.618 $\pm$ 0.029 & 14.022 $\pm$ 0.042 & 13.669 $\pm$ 0.033\\
70108 & 03140379+4703222 & 0.131 & 15.811 $\pm$ 0.080 & 15.183 $\pm$ 0.085 & 15.317 $\pm$ 0.141\\
10437 & 03150798+4709492 & 0.192 & 16.476 $\pm$ 0.136 & 15.894 $\pm$ 0.151 & 15.556 $\pm$ 0.188\\
00047 & 03152045+4706012 & 0.017 & 14.861 $\pm$ 0.036 & 14.504 $\pm$ 0.046 & 14.269 $\pm$ 0.051\\
20346 & 03152015+4719237 & 0.263 & 15.522 $\pm$ 0.054 & 15.184 $\pm$ 0.095 & 15.166 $\pm$ 0.112\\
20053 & 03151087+4715025 & 0.109 & 13.631 $\pm$ 0.021 & 13.062 $\pm$ 0.024 & 12.925 $\pm$ 0.030\\
10472 & 03145353+4712519 & 0.375 & 16.463 $\pm$ 0.125 & 16.133 $\pm$ 0.171 & 15.698 $\pm$ 0.190\\
20632 & 03152440+4719552 & 0.756 & 16.839 $\pm$ 0.175 & 15.903 $\pm$ 0.037 & 13.837 $\pm$ 0.039\\
20448 & 03153626+4717089 & 0.044 & 16.061 $\pm$ 0.074 & 15.448 $\pm$ 0.102 & 15.166 $\pm$ 0.103\\
20065 & 03150383+4714329 & 0.033 & 13.915 $\pm$ 0.022 & 13.377 $\pm$ 0.025 & 13.261 $\pm$ 0.029\\
50028 & 03134853+4716027 & 0.033 & 14.791 $\pm$ 0.036 & 14.170 $\pm$ 0.040 & 14.004 $\pm$ 0.050\\
20237 & 03145177+4719247 & 0.044 & 14.407 $\pm$ 0.036 & 13.652 $\pm$ 0.032 & 13.471 $\pm$ 0.035\\
20396 & 03145015+4715251 & 0.089 & 15.666 $\pm$ 0.067 & 14.892 $\pm$ 0.056 & 14.631 $\pm$ 0.071\\
60076 & 03135202+4708503 & 0.099 & 14.875 $\pm$ 0.036 & 14.507 $\pm$ 0.053 & 14.238 $\pm$ 0.060\\
00036 & 03152825+4704264 & 0.070 & 14.739 $\pm$ 0.056 & 14.132 $\pm$ 0.051 & 13.948 $\pm$ 0.051\\
50470 & 03142814+4716366 & 0.300 & 16.370 $\pm$ 0.121 & 15.750 $\pm$ 0.091 & 15.348 $\pm$ 0.158\\
60048 & 03142872+4710408 & 0.029 & 14.248 $\pm$ 0.029 & 13.589 $\pm$ 0.028 & 13.399 $\pm$ 0.035\\ \hline
20329 & 03152445+4718006 & 0.077 & 15.542 $\pm$ 0.054 & 14.841 $\pm$ 0.064 & 14.528 $\pm$ 0.063\\
40058 & 03135669+4723282 & 0.135 & 14.692 $\pm$ 0.037 & 14.046 $\pm$ 0.037 & 13.690 $\pm$ 0.041\\
\enddata										    
\end{deluxetable}

\section{VARIABLE CLASSIFICATION} \label{sec:vars}

None of the variables we have discovered and listed in Tables
\ref{tab:mvars} and \ref{tab:fvars} have been previously identified,
according to searches in the SIMBAD and VizieR online catalogs, and a
search of the relevant literature.  The large field of view, faint
magnitude limit, frequent time sampling, and the long observing
baseline of STEPSS provide a thorough and unequaled exploration of
the variable star content for this cluster.  STEPSS obtains 1\% and
10\% photometric precision from the saturation limit at $I\sim$15 down
to $I\sim$18 and $I\sim$20, respectively.  There are no other searches
in NGC 1245 with the time baselines and magnitude limits appropriate
for discovering the cluster variables.  For that reason, there are 
no previously known variables that we would have expected to find that were not found.

Here, we identify those variables for which we believe we have
sufficient information to classify as known variable types.  We use the 
variable classification methods described by \citet{sterken96}.

\subsection{Binaries} \label{sec:binaries}

Of the cluster members, objects 20176, 60193, 20534, 60303, and 10462 all have light curves characteristic 
of binary stars.  Objects 60303 and 10462 are detached eclipsing binaries, designated 
as EA variables, or Algol type binaries.  Objects 20176, 60193, and 20534 are 
contact binaries, designated as EW variables, or W UMa type binaries.  Of the field 
variables, objects 60017, 20065, 30143, 00276, 20398, 10437, 20274, and 10414 are all binary variables.

The periods and colors of contact binaries can be used as a rough
distance indicator.  A period-luminosity-color relation for contact
binaries was developed by \citet{rucin94}, and applied to contact
binaries discovered in the core of the cluster 47 Tucanae by
\citet{alb01}.

The relationship developed by Rucinski is:
\begin{equation} \label{equ:rucin}
M_V = -4.43 \,\, {\rm log} P + 3.63(V - I)_0 - 0.31 \,
\end{equation}
where $P$ is in days.  Using the derived $(V - I)_0$ colors (see \S\ref{sec:membership}) 
and the periods determined through the ANOVA algorithm, we calculate the distance 
modulus to the contact binary systems.  Given, the distance modulus for the cluster 
is $12.27 \pm 0.12$, the Rucinski formula gives us a distance modulus 
of $(m - M)_0 = 12.14$ for object 20176, $(m - M)_0 = 12.62$ for object 
60193, and $(m - M)_0 = 12.48$ for object 20534, consistent with the 
classification of these objects as cluster members.  The Rucinski relation has 
an intrinsic magnitude error of $\sim 0.3$, which is comparable to the systematic 
error found in \citet{alb01}.

\subsection{$\gamma$ Doradus Candidates} \label{sec:gamdor}

Early F-type stars containing
multiple frequencies with periods of 0.4 to 3 days and 0.01 to 0.1 mag
V-band amplitudes typify the $\gamma$ Doradus class \citep{KRI98,KAY99}.
The pulsation mechanism is thought to arise from high-order,
low-degree, nonradial gravity modes.  The light curves for stars 70025, 20196, 20223, and 20214 show
evidence of $\gamma$ Doradus variability.  We identify multiple frequencies
using the successive least-squares spectral analysis technique of
\citet{VAN69,VAN71} that is commonly applied to analyzing pulsating
variables \citep{HEN05}.  Applying the spectral analysis to all the
cluster variables, these four objects have unique signatures that set
them apart from the others.  For a majority of cluster variables, only
the fundamental and harmonics exist in the light curve, whereas for
the four $\gamma$ Doradus candidates, multiple frequencies exist that are
not integer fractions of the fundamental.

\begin{figure}
\epsscale{1.0}
\plotone{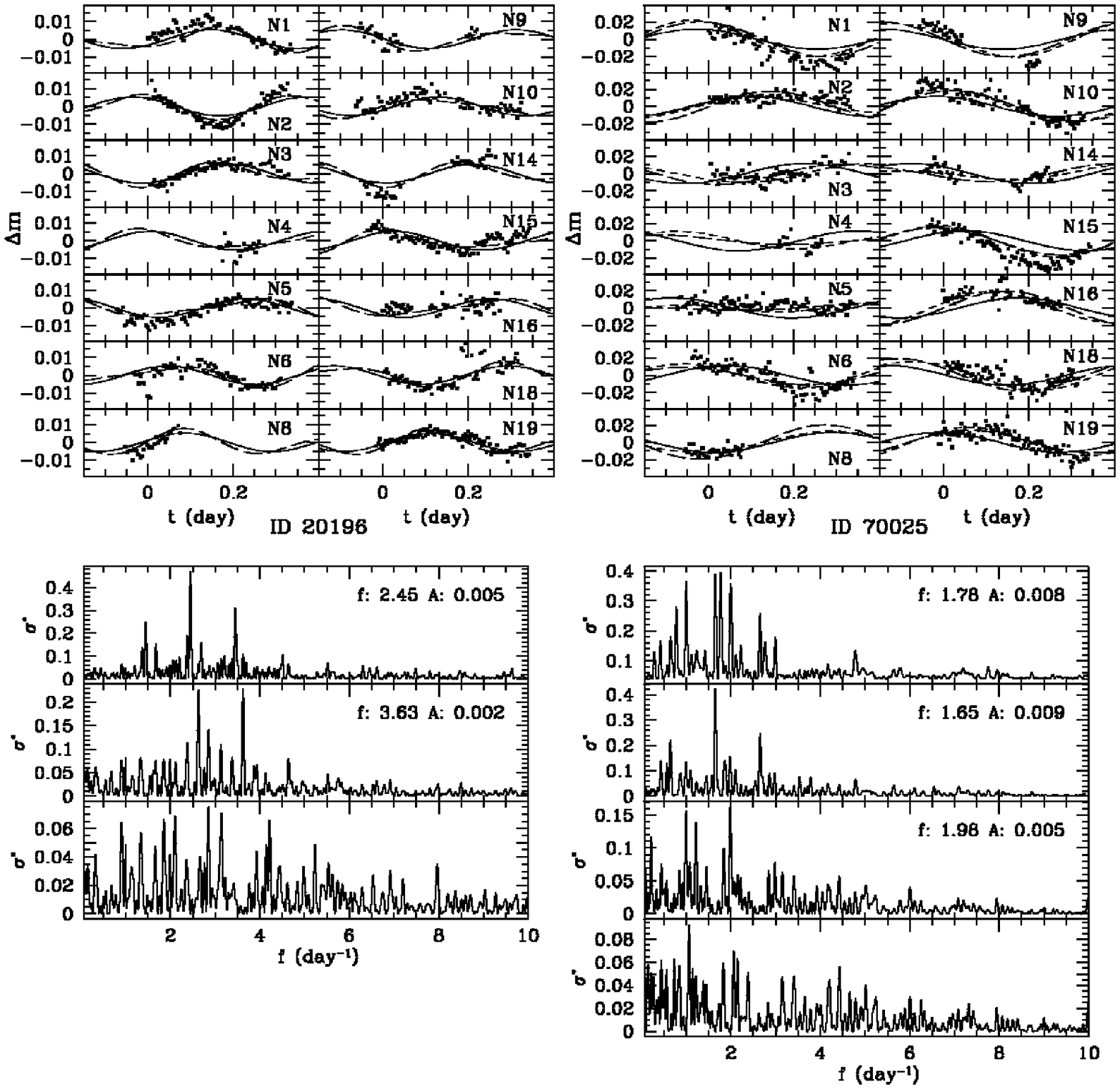}
\caption{Light curves and power spectra for $\gamma$ Doradus candidates 20196 and 
70025.  In the light curve plots in the upper left and upper right, each panel shows a 
single night's observations, with a label in the upper right corner of each panel identifying
the night of the observing run, and an arbitrary x-axis zero point.  The solid line 
shows a single component fit to the light curve, the long-dashed line gives a two-component 
fit, and the short-dashed line gives a three-component fit.
The lower left and lower right panels show successive optimum spectra for 20196 and 70025.
The top panel shows the spectra of the light curve which identifies
the strongest frequency component present in the data.  The
component's frequency and amplitude are labeled in the upper right
corner of the panel.  The successive panels from top to bottom show the spectra after taking
into account the known constituents identified in the previous panels.  The
bottom panel gives the spectra after taking into account the two
previously identified components.  No stable component is identified
in the bottom panel.}
\label{fig:gdor1}
\end{figure}

\begin{figure}
\epsscale{1.0}
\plotone{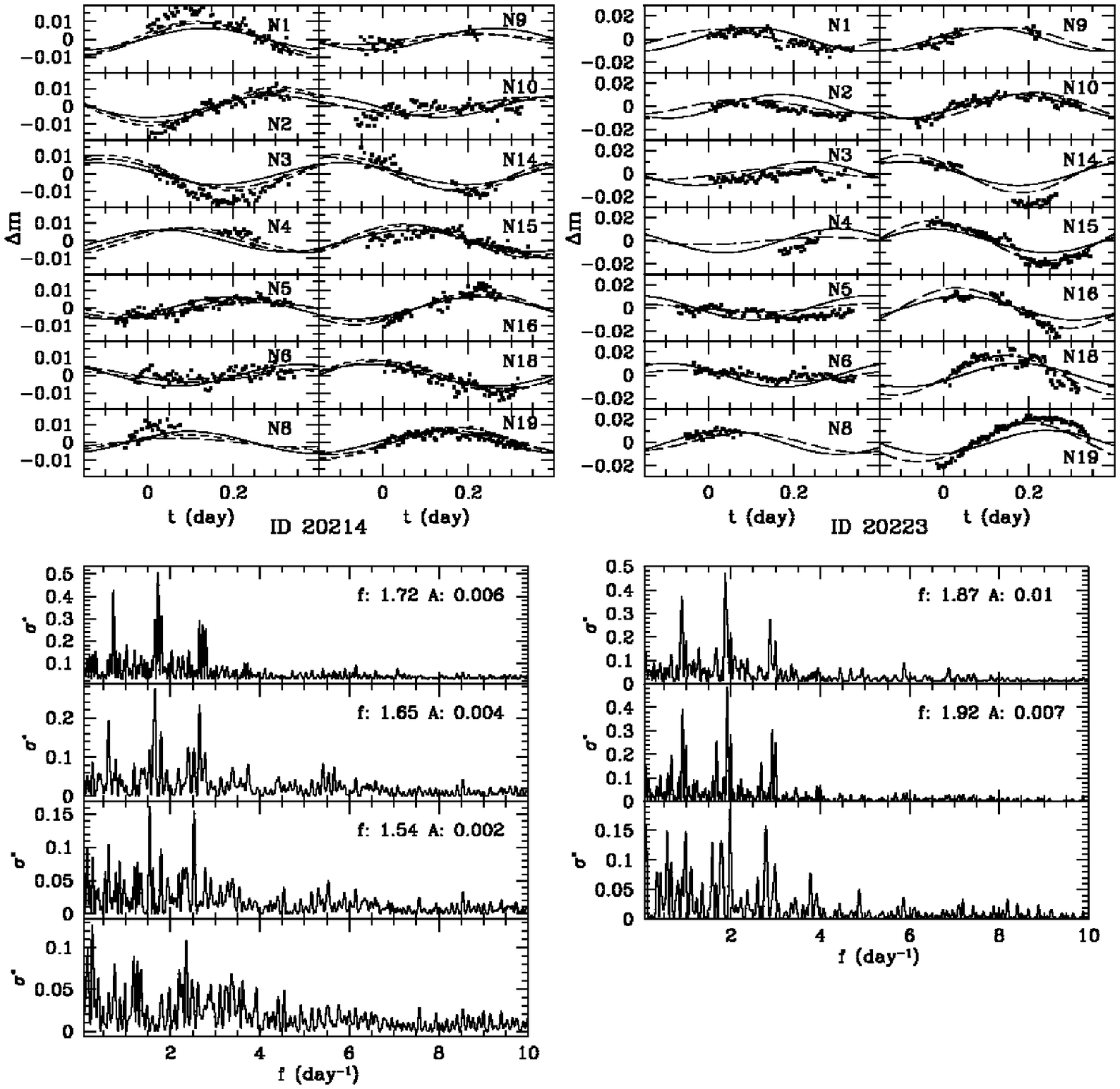}
\caption{Light curves and power spectra for $\gamma$ Doradus candidates 20214 and 
20223.  See caption of Figure \ref{fig:gdor1} for description. }
\label{fig:gdor2}
\end{figure}

Figures \ref{fig:gdor1} and \ref{fig:gdor2} present the light curves
for the $\gamma$ Doradus candidates along with the best-fit light
curve models in the upper parts of the figures.  Each panel shows a
single night's data and the lines represent models of the variability
with successively increased number of frequencies as identified in the
spectral analysis.  The solid line shows the first detected frequency
fit to the data, the long-dashed line gives a two-component fit, and
the short-dashed line gives a three-component fit.  Figures
\ref{fig:gdor1} and \ref{fig:gdor2} show the \citet{VAN71} optimum
spectra statistic of the light curve as a function of frequency in the
lower parts of the figures.  The panels show successive identification
of the frequencies present in the light curve from top to bottom.  If
a confident frequency is present, the frequency and its amplitude is
noted in the upper right hand corner of the panel.  We confidently
detect a frequency when it appears in all variants of the light curves
which we explain next.

When applying the spectral analysis to the cluster binary variables
60193 and 20534, their high amplitude, short period, and regularity
allows us to detect numerous frequencies which are all harmonics of the
fundamental.  However, in the original light curves, lower
significance frequencies were found very close to the dominant frequency
detected.  For binary stars this frequency splitting can occur if spot
patterns on the close binaries appear or vary.  The frequency
splitting can also occur if the exposure timing is not accurate, or 
as we discovered during this study, it can result from corrections
to the light curves to reduce systematic errors.

We find the procedure for removing the
systematic trend in a light curve performs
less than ideal in the presence of high amplitude variability.  It
slightly distorts the light curve enough to give rise to frequency
splitting.  The close frequency splitting is reduced when analyzing
light curves without the seeing trend correction.  In addition, the 
frequency splitting is further reduced by eliminating the last two nights 
of data.  This suggests that the timing may be an issue over the baseline 
of observations.  However, considering that the last two nights had some 
of the best seeing, it is not clear which of these two factors is the 
cause of the frequency splitting.  Overall, the
cleanest spectral analysis -- which detects only harmonics -- for
stars 60193 and 20534 occurs without the seeing correction and
without the last two nights of data.  Thus, we define confidently
detected frequencies in our $\gamma$ Doradus candidates as ones that are
stable and detected in all variants of the light curves: the original
light curves, light curves without the seeing correction, light curves
without the last two nights of data, and light curves without the seeing
correction and without the last two nights of data.

In addition to the multiple frequencies confidently identified in
these 4 objects that are not harmonically related, these objects also
occupy the upper main sequence in the CMD as expected for the typical
spectral type for $\gamma$ Doradus variables.  The large circles in
Figure~\ref{fig:cmd} designate their location in the CMD.  However, the
$\bv$ and estimated temperature for candidates 20214 and 20223
($\bv \sim 0.42$ and $T_{eff} \sim 6700$ K) are redward and cooler than the
empirically determined cool-edge boundary for $\gamma$ Doradus variables
as defined by \citet{HEN05} ($\bv$=0.38).  The empirical cool edge of
\citet{HEN05} agrees well with the theoretical $\gamma$ Doradus
instability strip as given by \citet{WAR03} ($T_{eff}=6850$ K).

Alternatively, our effective temperatures may be systematically
incorrect.  The effective temperatures for these objects are estimated
from the CMD and the best-fit cluster isochrone as described in
\citet{burke04}.  Varying the metallicity of the cluster between
-0.26$\leq [Fe/H] \leq$+0.13 while minimizing age, distance, and
reddening to yield a best-fit isochrone to the observed CMD, yields a
maximum $T_{eff}=6800$ K for these two stars.  Reaching this high
temperature requires an implausible $[Fe/H]=+0.13$.  Thus, if objects
20214 and 20223 are confirmed as $\gamma$ Doradus stars they may
represent the coolest known members of this class of variables.  Also,
the $\gamma$ Doradus candidates discovered in this study may represent
the oldest known members of this class of variables given the $\sim$1
Gyr age for NGC 1245 \citep{MAR02}.

Confirming the $\gamma$ Doradus variables will require follow up
spectroscopy and simultaneous light curves in several passbands.
However, alternative models for the variability cannot reproduce the
multiple frequencies that are not harmonically related.  This is
especially the case for objects 70025 and 20214 where we confidently
detect three components \citet{KRI98}.  Additionally, estimating the
physical parameters of the stars from the best-fit isochrone we
calculate the pulsation constant to range from -0.57$<\log Q<$-0.39
for our candidates.  These values are consistent with the confirmed
$\gamma$ Doradus variables of \citet{HAN02}.  We used the strongest,
first frequency detected when calculating the pulsation constant.

Assuming these four objects are $\gamma$ Doradus variables, we are able 
to measure the fraction of cluster members exhibiting this type of 
variability.  \citet{burke04} statistically determine the cluster membership by
characterizing the background star counts as a function of apparent
magnitude from a control field at the outskirts of the field of view.
The cluster membership as a function of magnitude is shown as Figure 9
in \citet{burke05}.  There is an estimated 45.9 cluster members for NGC
1245 between 15.83$<V<$16.27, the apparent magnitudes of our brightest
and faintest $\gamma$ Doradus candidates.  This corresponds to
2.88$<(M_{V})_{o}<$3.32 assuming a cluster distance modulus
$(m-M)_{o}=12.27$ and extinction $A_{V}=0.68$.  If all four objects
are confirmed as $\gamma$ Doradus variables, then 8.7\% of stars with
properties similar to NGC 1245 in the above absolute magnitude range
exhibit $\gamma$ Doradus variability.  If none of the objects end up
being $\gamma$ Doradus variables, then a null result implies $<6.5\%$
(95\% confidence) of stars in this magnitude range exhibit $\gamma$
Doradus variability assuming we are 100\% complete in detecting
variability with amplitudes similar to the candidates.

\section{CONCLUSIONS}

We have discovered 43 previously unknown variables in the field of NGC
1245, of which 14 are potential cluster members.  Many of these
variables are very low amplitude.  Our method for determining relative
photometry allows us to achieve sub-1\% precision for stars brighter
than $V=18$, enabling the characterization of such low amplitude
variables.  The techniques described in this paper can be readily
applied to other long-term photometric data sets to determine the
variability content of other surveys.

We roughly characterize the variables potentially belonging to the
cluster.  The light curves for objects 20534, 20176, 60193, 10462, and
60303 contain strong evidence for binarity.  Object 10462 and
60303 are detached eclipsing binaries.  If their membership in this 
cluster is confirmed, then the known ages and composition of these binaries 
offer opportunities for testing stellar models for cool stars. 

The light curves for
objects 20196, 70025, 20223, 20214 have evidence of multiple
frequencies present and located near the main sequence turnoff.  These
objects are possible $\gamma$ Doradus candidates.  Objects 20223 and
20214 would be the coolest ($T_{eff} \sim 6700$ K) known $\gamma$ Doradus
variables if their status is verified.  The standard solar metallicity
and solar calibrated mixing length models of $\gamma$ Doradus
variability do not predict pulsations in stars this cool
\citep{DUP04,WAR03}.  Further exploration of parameter space may be
necessary to fit these cool $\gamma$ Doradus candidates.  Alternatively, the variability
for these four objects can result from rapid rotation coupled with
unstable spot patterns on the surface.  Measuring the rotation rates
and simultaneous light curves in multiple passbands for these objects
would settle their classification.  

The nature of potential cluster
members 00210, 30338, 10429, 20513, and 20510 remain uncertain.  These
objects most likely result from ellipsoidal variability.
Of the non-cluster objects, 10414, 30143, 20274, 20398, 10437, and
20065 are eclipsing binaries.  The rest of the non-members cannot be
well-characterized without knowledge of their absolute brightnesses
and colors.

\acknowledgments

We would like to thank Marc Pinsonneault, Darren DePoy, Kris Stanek, and Rick Pogge for helpful comments.  We 
thank Christopher Alard for the implementation of the ANOVA period analysis.  This 
work was supported by the National Aeronautics and Space Administration under Grant No. NNG04GO70G issued 
through the Origins of Solar Systems program.

\end{document}